%% file: main.tex
\documentclass[10pt,sigconf]{acmart}

\usepackage[T1]{fontenc}
\usepackage{tabularx}
\usepackage{subfigure}
\usepackage{wrapfig}
\usepackage{pbox}
\usepackage{bbding}
\usepackage{textcomp}
\usepackage{xcolor}
\usepackage[noend]{algorithm2e}
\usepackage{bbm}
\usepackage{dblfloatfix}
\usepackage{pifont}
\usepackage{bm}
\usepackage{enumitem}

\ifodd 1
\newcommand{\frev}[1]{#1}

\ifodd 0
\newcommand{\rev}[1]{{\color{blue}#1}}      
    
\newcommand{\needrev}[1]{{\color{green}#1}} 
\else
\newcommand{\rev}[1]{#1} 
 
\newcommand{\needrev}[1]{#1} 
\fi

\def\BibTeX{{\rm B\kern-.05em{\sc i\kern-.025em b}\kern-.08em
    T\kern-.1667em\lower.7ex\hbox{E}\kern-.125emX}}

\setcopyright{acmcopyright}
\acmYear{2023}
\copyrightyear{2023}
\setcopyright{rightsretained}
\acmConference[ACM MobiCom'23]{The 29th Annual International Conference on Mobile Computing and Networking}{October 2--6, 2023}{Madrid, Spain}
\acmBooktitle{The 29th Annual International Conference on Mobile Computing and Networking (ACM MobiCom'23), October 2--6, 2023, Madrid, Spain}\acmDOI{10.1145/3570361.3613290}
\acmISBN{978-1-4503-9990-6/23/10}

\input{defines.tex}
\input{glossaries.tex}

\begin{document}

\newcommand{\name}{MUSE-Fi\xspace}
\title{\name: Contactless MUti-person SEnsing Exploiting Near-field Wi-Fi Channel Variation}

\ifodd 0
\author{
	Conditional Accepted Paper \#902 to ACM MobiCom 2023
    }
    \thanks{*~***. \vspace{-1ex}}
\affiliation{
    \institution{\color{white}
	$^1$ABA \country{AA} \\ \vspace{-.5ex}
	$^2$BAB \country{BB} \\
    Email: aaa@bbb.cc
        }
    }
    \renewcommand{\authors}{Conditionally Accepted Paper \#902 to ACM MobiCom 2023}
    \renewcommand{\shortauthors}{Conditionally Accepted Paper \#902 to ACM MobiCom 2023}
\else
\author{ {\Large
    Jingzhi Hu$^{1*}$\quad Tianyue Zheng$^{1*}$\quad Zhe Chen$^{2}$\quad Hongbo Wang$^{1}$\quad Jun Luo$^{1}$
    }}
    \thanks{*~Both authors contributed equally to this research. \vspace{-1ex}}
\affiliation{
   \institution{{\normalsize
	$^1$School of Computer Science and Engineering, Nanyang Technological University (NTU) \country{Singapore} \\
	$^2$Intelligent Networking and Computing Research Center and School of Computer Science, Fudan University \country{China} \\
    Email: \{jingzhi.hu, tianyue002, hongbo001, junluo\}@ntu.edu.sg, zhechen@fudan.edu.cn
        }}
    }
    \renewcommand{\authors}{J. Hu, T. Zheng, Z. Chen, H, Wang, and J. Luo}
    \renewcommand{\shortauthors}{J. Hu, T. Zheng, Z. Chen, H, Wang, and J. Luo}
\fi

\begin{abstract}
Having been studied for more than a decade, Wi-Fi human sensing still faces a major challenge in the presence of multiple persons, simply because the limited bandwidth of Wi-Fi fails to provide a sufficient range resolution to physically separate multiple subjects. Existing solutions mostly avoid this challenge by switching to radars with GHz bandwidth, at the cost of cumbersome deployments. Therefore, \textit{could Wi-Fi human sensing handle multiple subjects} remains an open question. 
This paper presents \name, the first Wi-Fi multi-person sensing system \frev{with physical separability}.
The principle behind \name is that, given a Wi-Fi device (e.g., smartphone) very close to a subject, the \textit{near-field} channel variation caused by the subject significantly overwhelms variations caused by other distant subjects. Consequently, focusing on the channel state information (CSI) carried by the \rev{traffic} in and out \rev{of} this device naturally allows for physically separating multiple subjects. Based on this principle, we propose three sensing strategies for \name: i) uplink CSI, ii) downlink CSI, and iii) downlink beamforming feedback, where we specifically tackle signal recovery from sparse (per-user) \rev{traffic} under realistic multi-user communication scenarios. Our extensive evaluations clearly demonstrate that \name is able to successfully handle multi-person sensing with \rev{respect} to three typical applications: respiration monitoring, gesture detection, and activity recognition.

\end{abstract}




\maketitle
\vspace{-0.5em}
\section{Introduction}\label{sec:introduction} 
\input{1_introduction}

\section{Sensing by Near-field Domination}  \label{sec:nfd}
\input{2_nearfield}

\section{Shaping-up \name}\label{sec:design} 
\input{3_system_design}

\section{Prototyping \& Experiment Setup} \label{sec:implementation}
\input{4_implementation}

\section{Evaluations}  \label{sec:evaluation}
\input{5_evaluations}

\vspace{-.33em}
\section{Related Work}\label{sec:related_work}
\input{6_related_work}

\vspace{-.2em}
\section{Conclusion}\label{sec:conclusion}
\input{8_conclusion}

\balance

\newpage
\bibliographystyle{ACM-Reference-Format}
\bibliography{reference}

\end{document}

%% file: defines.tex
\newcommand{\iu}{\mathrm{i}}

\newcommand{\asin}{\mathrm{arcsin}}
\newcommand{\diag}{\bm{\mathrm{diag}}}
 

%% file: glossaries.tex
\usepackage{glossaries}
\usepackage[symbols,nogroupskip]{glossaries-extra}

\newglossaryentry{tar}
{
	name={\ensuremath{\mathcal S}},
	sort={p},
	description={Subscript of Target User}
}

\newglossaryentry{int}
{
	name={\ensuremath{\mathcal I}},
	sort={p},
	description={Subscript of Interfering User}
}

\newglossaryentry{int_i}
{
	name={\ensuremath{\mathcal I_i}},
	sort={p},
	description={Subscript of Interfering User}
}

\newglossaryentry{int_j}
{
	name={\ensuremath{\mathcal I_j}},
	sort={p},
	description={Subscript of Interfering User}
}

\newglossaryentry{num_ue_ant}
{
	name={\ensuremath{A_{\mathrm{ue}}}},
	sort={p},
	description={Number of UE's antenna}
}

\newglossaryentry{num_ap_ant}
{
	name={\ensuremath{A_{\mathrm{ap}}}},
	sort={p},
	description={Number of AP's antenna}
}

\newglossaryentry{numAddr}
{
	name={\ensuremath{N_{\mathrm{mac}}}},
	sort={p},
	description={Number of AP's antenna}
}

\newglossaryentry{numInterferer}
{
	name={\ensuremath{N_{\gls{int}}}},
	sort={p},
	description={Number of AP's antenna}
}

\newglossaryentry{setInterferer}
{
	name={\ensuremath{\{1,... N-1\}}},
	sort={p},
	description={Number of AP's antenna}
}

\newglossaryentry{threshold}
{
	name={\ensuremath{\gamma_{\mathrm{th}}}},
	sort={p},
	description={Number of AP's antenna}
}

\newglossaryentry{numSubCarrier}
{
	name={\ensuremath{N_{\mathrm{SC}}}},
	sort={p},
	description={Number of subcarriers}
}

\newglossaryentry{numSTS}
{
	name={\ensuremath{N_{\mathrm{STS}}}},
	sort={p},
	description={Number of space time stream}
}

\newglossaryentry{numPhase}
{
	name={\ensuremath{N_{\mathrm{P}}}},
	sort={p},
	description={Number of Rx antennas}
}

\newglossaryentry{numFragments}
{
	name={\ensuremath{N_{\mathrm{F}}}},
	sort={p},
	description={Number of Rx antennas}
}

\newglossaryentry{rsampleFreq}
{
	name={\ensuremath{f_{\mathrm{rs}}}},
	sort={p},
	description={Frequency of resampling}
}

\newglossaryentry{cutoffFreq}
{
	name={\ensuremath{f_{\mathrm{cut}}}},
	sort={p},
	description={Frequency of resampling}
}

\newglossaryentry{nspThreshold_N}
{
	name={\ensuremath{N_{\mathrm{nsp}}}},
	sort={p},
	description={Frequency of resampling}
}

\newglossaryentry{nspThreshold_t}
{
	name={\ensuremath{\Delta t}},
	sort={p},
	description={Frequency of resampling}
}

\newglossaryentry{nspThreshold_T}
{
	name={\ensuremath{T_{\mathrm{t}}}},
	sort={p},
	description={Frequency of resampling}
}

\newglossaryentry{num_nsp_slice}
{
	name={\ensuremath{F_{\mathrm{nsp}}}},
	sort={p},
	description={Number of non-sparse time slices}
}

\newglossaryentry{num_sp_slice}
{
	name={\ensuremath{F_{\mathrm{sp}}}},
	sort={p},
	description={Number of sparse time slices}
}

\newglossaryentry{freq_resample}
{
	name={\ensuremath{f_{\mathrm{rs}}}},
	sort={p},
	description={Number of sparse time slices}
}

\newglossaryentry{numHid}
{
	name={\ensuremath{H}},
	sort={p},
	description={Number of hidden layers}
}

\newglossaryentry{filterDim}
{
	name={\ensuremath{L_{\mathrm{kn}}}},
	sort={p},
	description={size of kernel filter}
}

\newglossaryentry{numResBlock}
{
	name={\ensuremath{L_{\mathrm{f}}}},
	sort={p},
	description={size of kernel filter}
}

\newglossaryentry{numResuse}
{
	name={\ensuremath{N_{\mathrm{re}}}},
	sort={p},
	description={size of kernel filter}
}

\newglossaryentry{numChannel}
{
	name={\ensuremath{N_{\mathrm{ch}}}},
	sort={p},
	description={size of channel}
}

\newglossaryentry{numFreqCp}
{
	name={\ensuremath{N_{\mathrm{F}}}},
	sort={p},
	description={number of frequency components}
}

\newglossaryentry{numSample}
{
	name={\ensuremath{N_{\mathrm{s}}}},
	sort={p},
	description={size of channel}
}

\newglossaryentry{numTx}
{
	name={\ensuremath{N_{\mathrm{tx}}}},
	sort={p},
	description={Number of UE's antenna}
}
\newglossaryentry{numRx}
{
	name={\ensuremath{N_{\mathrm{rx}}}},
	sort={p},
	description={Number of UE's antenna}
}

%% file: 1_introduction.tex
Since we were first able to obtain CSI (channel state information) in certain Wi-Fi devices~\cite{CSI-CCR11}, \textit{Wi-Fi human sensing}~\cite{LiFS-MobiCom16,WiAG-MobiSys17,Widar2-MobiSys18,Widar3-MobiSys19,WiPose-MobiCom20,Wang2022Placement,VitalSign-MobiHoc15,MultiSense-UbiComp20} has been attracting significant \rev{attention} from both academia and industry. During the past decade or so, many applications of Wi-Fi human sensing have been developed, notably including vital signs monitoring~\cite{VitalSign-MobiHoc15,MultiSense-UbiComp20}, gesture detection~\cite{Widar3-MobiSys19, WiAG-MobiSys17}, activity recognition~\cite{WiPose-MobiCom20, RF-Net-SenSys20}, as well as localization and motion tracking~\cite{ilocScan, Widar2-MobiSys18, LiFS-MobiCom16, chen2019m}. Whereas such sensing applications have a promising potential to be integrated with the ubiquitously deployed Wi-Fi communication infrastructures, they all face a major obstacle in conducting realistic \textit{multi-person sensing: the limited Wi-Fi bandwidth fails to offer a sufficient range resolution to distinguish different sensing subjects}.

Because Wi-Fi communication does not seem to embrace a super-wide bandwidth due to its contention-based multi-access nature,\footnote{The 320~\!MHz bandwidth of Wi-Fi~7~\cite{khorov2020current} only yield a meter-level range resolution, yet leading to insufficient sampling rates due to frame aggregation.} existing sensing proposals often avoid its limitation by resorting to radars with a GHz-level bandwidth~\cite{adib2015smart, PhySep-SenSys22}, yet radar sensing is somehow inferior to Wi-Fi sensing as it demands extra deployments. In order to continue exploiting Wi-Fi's potential in integrated sensing and communication~(ISAC)~\cite{isacot,isac_wifi}, two makeshifts are often adopted. 
On one hand, many distributed antennas can be used to achieve enhanced spatial resolution for separating subjects~\cite{Widar2-MobiSys18}, at the cost of messing up with the Wi-Fi communication functions. 
\frev{On the other hand, signal processing techniques for separating five subjects at the CSI level have been attempted~\cite{MultiSense-UbiComp20} without offering guaranteed separability in general~\cite{zhang2022can, PhySep-SenSys22}.}

\begin{figure}[t]
    \setlength\abovecaptionskip{6pt}
    \centering
    \includegraphics[width=.83\linewidth]{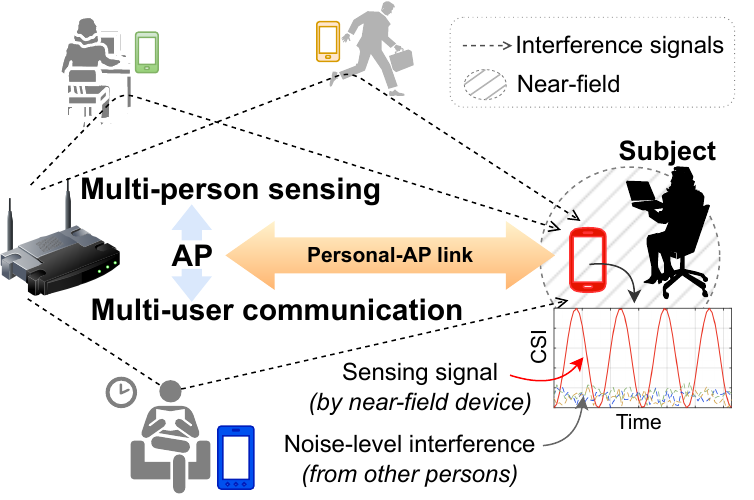}
    \caption{\frev{While each personal device \textit{uniquely} identifies a person, the sensing signal (upon the person) offered by the identifying device within \textit{near-field} overwhelms the interference from other persons.}}
    \label{fig:overview}
    \vspace{-3ex}
\end{figure}
In reality (as in Figure~\ref{fig:overview}), each person often has its own wearable Wi-Fi devices, typically a smartphone or even a smartwatch. 
Although the communication link between such a personal device and the nearby Wi-Fi access point (AP) is deemed as the basic sensing media by earlier proposals on Wi-Fi human sensing, those proposals aim to leverage either a single link to perform sensing~\cite{Wang2022Placement,VitalSign-MobiHoc15,Widar2-MobiSys18,MultiSense-UbiComp20} or multiple links to offer a slightly improved spatial resolution~\cite{LiFS-MobiCom16,Widar-MobiHoc17,Widar3-MobiSys19,WiPose-MobiCom20}. %WiAG-MobiSys17 
They all neglect two fundamental factors in such a realistic multi-user communication setting shown in Figure~\ref{fig:overview}: 
i) each personal-AP link uniquely identifies the human subject to be sensed, and 
ii) since the subject is within the \textit{near-field} (less than 0.2~\!m in range) of its own Wi-Fi device, the channel variation caused by its motions to its personal-AP link could be so strong as to push the interference from other subjects down to the noise floor. 
In other words, the default multi-user communication setting of Wi-Fi does offer the potential to be naturally extended to multi-person sensing, if one can properly integrate sensing into communication.

From application perspective, such near-field multi-person sensing naturally supports various functions under the pervasive deployment of Wi-Fi infrastructure. As these functions include sensing vital signals, gestures, activities, and locations, they are especially applicable to eXtended Reality~(XR).
In particular, integrating gestures and activities recognition into Wi-Fi communication reduces the peripheral sensors, leading to lighter and less power-consuming virtual reality~(VR) and merged reality~(MR) headsets, making them more desirable for long-time wearing~\cite{Akyildiz2022Wireless,Tan2022Commodity}.
Furthermore, the environmental and human sensing results indicate key contextual and localization information of nearby human and object motions; 
overlaying such information on the top of real-world visions facilitates augmented reality~(AR) and MR applications in intrusion detection, patient monitoring, and machine status assessing~\cite{Carmigniani2011Augmented, Li2020Wireless}.

Nonetheless, integrating multi-person sensing with multi-user communication is highly non-trivial, as existing practices, exploiting only artificial Wi-Fi \rev{traffic} for the sensing purpose, barely offer any experience. 
In practice, multi-user scenarios typically cause a much lower and very irregular frame arrival rate per link, thanks to the contention-based multi-access nature of Wi-Fi. 
Since the CSI carried by each frame is a critical channel state sample for Wi-Fi sensing, a lower and irregular frame rate indicates a lower and irregular sampling rate, which may significantly confine the usability of Wi-Fi sensing. As most Wi-Fi sensing applications have been developed upon a high and regular frame rate (up to 1000~\! frame/s~\cite{MultiSense-UbiComp20,Widar2-MobiSys18,WiPose-MobiCom20}), this challenge, crucial to seamless integration of multi-person sensing with multi-user communication for Wi-Fi, has never been seriously tackled.

To address these challenges, we propose \name as a novel \underline{MU}lti-person \underline{SE}nsing system leveraging Wi-\underline{Fi}. 
To motivate
 \name, we first theoretically characterize and experimentally verify the \textit{dominating effect} in near-field Wi-Fi sensing, upon which we develop criteria on the physical separability of multiple subjects. 
Based on the theoretical characterizations, we propose three sensing strategies for \name to be integrated with the \rev{traffic} cross each personal-AP link, namely exploiting i) uplink (to AP) CSI, ii) downlink (from AP) CSI, and iii) downlink BFI (beamforming feedback information)~\cite{bejarano2013ieee}.
For all strategies, we propose a \emph{sparse recovery algorithm}~(SRA) to mask the potential variance in frame rate; it aims to regulate the input samples so as to deliver a unified data flow to later processing pipelines for respective sensing functions. 
In addition, we study the sensing effectiveness of these strategies by contrasting the BFI-enabled compressive sensing with conventional CSI-based Wi-Fi sensing. Our key contributions can be summarized as follows:
\begin{itemize}[leftmargin=*]
    \item We propose \name as the first true multi-person Wi-Fi sensing system; it integrates multi-person sensing with multi-user communication in a seamless manner.
    \item We, for the first time, expose the dominating effect of near-field Wi-Fi sensing; it is exploited by \name to achieve physical separation of multiple subjects.
    \item We design three sensing strategies for \name and equip them with an SRA to mask the variance in frame rate.
    \item We reveal the pros and cons of BFI-enabled Wi-Fi sensing against the conventional CSI-enabled one.
    \item We implement \name prototype and evaluate it with extensive experiments. The promising results confirm that \name indeed supports multi-person Wi-Fi sensing under realistic scenarios.
\end{itemize}

The rest of the paper is organized as follows. Section~\ref{sec:nfd} discusses the dominating effect of near-field sensing both theoretically and experimentally. 
Section~\ref{sec:design} presents the sensing strategies for \name, along with the crucial SRA to regulate the frame rate. Section~\ref{sec:implementation} specifies how the \name prototype is implemented and how the application scenarios for case studies are set up. Section~\ref{sec:evaluation} reports the evaluation results \rev{of} three case studies. Related works are briefly discussed in Section~\ref{sec:related_work}.
Finally, Section~\ref{sec:conclusion} concludes our paper. 

%% file: 2_nearfield.tex
In this section, we introduce the Wi-Fi human sensing basics, and systematically study and validate the dominating signal variations in near-field sensing.
\frev{Compared with conventional antenna near-field and capacitive coupling~\cite{Balanis2016Antenna,Grosse2014Capacitive} not developed for practical multi-person sensing, our theoretical analyses allow for characterizing the feasible region of near-field sensing and shedding insights on the upper/lower bounds of subject number and spacing.}

\subsection{Wi-Fi Human Sensing Basics} 
\label{ssec:basics}

We start \rev{by} introducing a Wi-Fi sensing system with \rev{an AP and \textit{user equipment}~(UE) pair} aiming to sense the physical motion of a human \emph{subject} denoted by \gls{tar}.
At time $t$, denote \rev{the} distance between the AP and $\gls{tar}$ by $d_{A, \gls{tar}}(t)$ and the distance between $\gls{tar}$ and the UE by $d_{\gls{tar}, E}(t)$.
Further focusing on the influence of \gls{tar}, we model the wireless channel gain between the AP and the UE as:
\begin{align}
\label{equ: overall channel}
h_{A,E}(t) =  h_{A,\gls{tar},E}(t) +h_{A,E}^{S} + h_{A,E}^{D}(t), 
\end{align}
where $h_{A,E}^{S}$ and $h_{A,E}^{D}(t)$ represent the static and dynamic channel gains between the AP and UE due to, respectively, the direct communication path and interfering motions along it, and $h_{A,\gls{tar},E}(t)$ indicates the channel gain from the AP to UE via the reflection of $\gls{tar}$, which can be expressed as:
\begin{equation}
\label{equ: h_ij target}
\!\!\!h_{A,\gls{tar},E}(t) = \frac{\lambda^2\sqrt{G_{A, \gls{tar}, E}}\exp(-\iu{2\pi}(d_{A,\gls{tar}}(t)+d_{\gls{tar},E}(t))/{\lambda})}{(4\pi)^2 (d_{A,\gls{tar}}(t)d_{\gls{tar},E}(t))^{\alpha/2}},
\end{equation}
where $\lambda$ is the carrier wavelength, $G_{A, \gls{tar}, E}$ represents the product of Tx and Rx antenna gains and the reflection coefficient of \gls{tar}, and $\alpha$ is the path loss exponent~\cite{goldsmith2005wireless}.
Typically, $\alpha \in [2, 4]$ with 
$\alpha\approx 4$ for indoor environments~\cite{rappaport2010wireless}.
Therefore, Wi-Fi human sensing can be described as follows: the physical motion of a human subject results in changes of $d_{A,\gls{tar}}(t)$ and~$d_{\gls{tar},E}(t)$, which in turn lead to the changes of channel gain $h_{A,\gls{tar},E}(t)$ over time.
Therefore, by analyzing the time series of $h_{A,\gls{tar},E}(t)$ obtained from the CSI of the Wi-Fi frames, both AP and UE are able to sense the motion of \gls{tar}.

\subsection{Feasible Region for Near-field Sensing}
\label{s2ec: near-field domination}

Consider a more general scenario where two persons exist in the Wi-Fi sensing system.
Without loss of generality, we let one of them be the subject \gls{tar}, and refer to the other as the \emph{interferer} denoted by \gls{int}.
Consequently, the channel gain between the AP and UE in Eqn.~\eqref{equ: overall channel} becomes:
\begin{align}
\label{equ: overall channel with interference}
\tilde{h}_{A,E}(t) =  h_{A,\gls{tar},E}(t) + h_{A,\gls{int},E}(t) +h_{A,E}^{S} + h_{A,E}^{D}(t),
\end{align}
where $h_{A,\gls{int},E}(t)$ is the channel gain from the AP to UE via the reflection of $\gls{int}$; it can be modeled in a similar manner as Eqn.~\eqref{equ: h_ij target}.
Eqn.~\eqref{equ: overall channel with interference} seems to suggest that it is hard to separate the channel influences imposed by \gls{tar} and \gls{int} since their channel gains get mixed up.
Nevertheless, we point out that, in the Wi-Fi sensing scenarios where \gls{tar} is close to or in the near-field of UE~(i.e., distance below 0.2~\!m, empirically), the variation of the channel gain is dominated by the \gls{tar}'s physical motion; in other words, $\partial |h_{A,\gls{tar},E}(t)|/\partial t \gg \partial |h_{A,\gls{int},E}(t)|/\partial t$.
We term this phenomenon \textbf{near-field domination} effect, and we provide its theoretical analysis as follows.

Firstly, to quantify the variation of $h_{A,\gls{tar},E}(t)$, we evaluate it by the squared amplitude of the partial derivative of $h_{A,\gls{tar},E}(t)$ w.r.t. $t$, which is referred to as the \emph{power of channel variation}.
To simplify the analysis, we assume $\partial d_{A,\gls{tar}}(t)/\partial t = \partial d_{\gls{tar}, E}(t)/\partial t = v_{\gls{tar}}$.
The value of $v_{\gls{tar}}$ can be interpreted as the \textit{intensity} of \gls{tar}'s motion in terms of speed.
The power of channel variation of \gls{tar} can then be calculated as:
\begin{align}
\label{equ: csi variation power}
P_{\gls{tar}} & = \left|\frac{\partial h_{A,\gls{tar},E}(t)}{\partial t} \right|^2 \nonumber \\[-.2em]
& \approx  \frac{{G_{A,\gls{tar},E}}\lambda^4 v_{\gls{tar}}^2}{(4\pi)^4(d_{A,\gls{tar}}d_{\gls{tar},E})^\alpha}
\left[\frac{\alpha^2}{4}\Big(\frac{d_{A,\gls{tar}} + d_{\gls{tar},E}}{d_{A,\gls{tar}}d_{\gls{tar},E}}\Big)^2 + \frac{16\pi^2}{\lambda^2}\right]\nonumber\\[-.3em]
& \stackrel{(\star)}{\approx}{\tilde{G}_{A,\gls{tar},E}}\cdot v_{\gls{tar}}^2\cdot {(d_{A,\gls{tar}}d_{\gls{tar},E})^{-\alpha}},
\end{align}
where we omit symbol $t$ in the distance notations and let $\tilde{G}_{A,\gls{tar},E} = (\lambda/4\pi)^2 G_{A,\gls{tar},E}$ for the sake of brevity.
In~the second row of Eqn.~\eqref{equ: csi variation power}, the first term inside the bracket is caused by the amplitude variation of the channel gain and the second term results from the phase variation of the channel gain.
$(\star)$ holds because, in typical 5~\!GHz Wi-Fi sensing systems with \gls{tar} in the near-field of UE (e.g., $d_{A,\gls{tar}}\!\sim\!3$~\!m, $d_{\gls{tar},E}\!\sim\! 0.1$~\!m, and $\lambda\!\sim\! 0.06$~\!m), the first term in the bracket is much smaller than the second term and thus can be omitted, implying that the channel variation is mainly due to that \gls{tar}'s motion changes the phase of the channel.

The power of variation of $h_{A,\gls{int},E}(t)$ can be similarly derived for \gls{int} as $P_{\gls{int}} = \tilde{G}_{A,\gls{int},E}v_{\gls{int}}^2(d_{A,\gls{int}}d_{\gls{int},E})^{-\alpha}$, with $d_{A,\gls{int}}$ and $d_{\gls{int},E}$ being the distance between the AP and \gls{int} and between \gls{int} and the UE, respectively, and $v_{\gls{int}}$ being the intensity.
Consequently, the near field domination effect can be interpreted as $P_{\gls{tar}}$ being significantly larger than $P_{\gls{int}}$, thanks to $d_{\gls{tar},E}$ being much smaller than $d_{\gls{int},E}$, given $d_{A,\gls{tar}}\approx d_{A,\gls{int}}$ and $v_{\gls{tar}}\approx v_{\gls{int}}$. 
It is worth noting that assuming $v_{\gls{tar}}\approx v_{\gls{int}}$ may not be practical, because human sensing to different targets may have distinct meaning (e.g., respiration monitoring against gesture detection). 
Though our following derivation shall stick to this symmetric assumption for the ease of exposition, we will experimentally validate that the near-field domination still holds even with asymmetric intensities, as far as there is a sufficient discrepancy between $d_{\gls{tar},E}$ and $d_{\gls{int},E}$.

In order to concretely characterize the interferer's feasible region that maintains the domination of $P_{\gls{tar}}$ at the UE, we propose an novel metric \textbf{variation to interference ratio}~(VIR); it evaluates the variation power ratio between $h_{A,\gls{tar},E}(t)$ and the sum of $h_{A,\gls{int},E}(t)$ and dynamic channel $h_{A,E}^{D}(t)$. 
Based on~\cite{Wang2022Placement}, $h_{A,E}^{D}(t)$ can be also treated as an interference, whose power $P_d$ is in linear proportion to that of the static channel gain.
Therefore, assuming a LoS path between the AP and UE, we have $P_{d} = \eta {\lambda}^2 {d_{A,E}}^{-\alpha} + b$, where $\eta$ and $b$ are fixed for a given pair of AP and UE. 
Then, we have:
\begin{align}
\label{equ: vsir}
\!\!\!\mathrm{VIR}_{\gls{tar}} \!=\! \frac{P_{\gls{tar}}}{P_{\gls{int}} \!+\! P_d} 
\!=\! \frac{v_{\gls{tar}}^2{\tilde{G}_{A, \gls{tar}, E}}{(d_{A,\gls{tar}}d_{\gls{tar},E})^{-\alpha}}}{\eta \lambda^2 d_{A,E}^{-\alpha} + b+v_{\gls{int}}^2\tilde{G}_{A,\gls{int},E}(d_{A,\gls{int}}d_{\gls{int},E})^{-\alpha}}.\!\!
\end{align}
Intuitively, the feasible region of \gls{int} is indicated by $\mathrm{VIR}_{\gls{tar}}$ value being greater than a threshold \gls{threshold}. 

\begin{figure}[b]
    \vspace{-2ex}
    \setlength\abovecaptionskip{3pt}
    \centering
    \includegraphics[width=.73\linewidth]{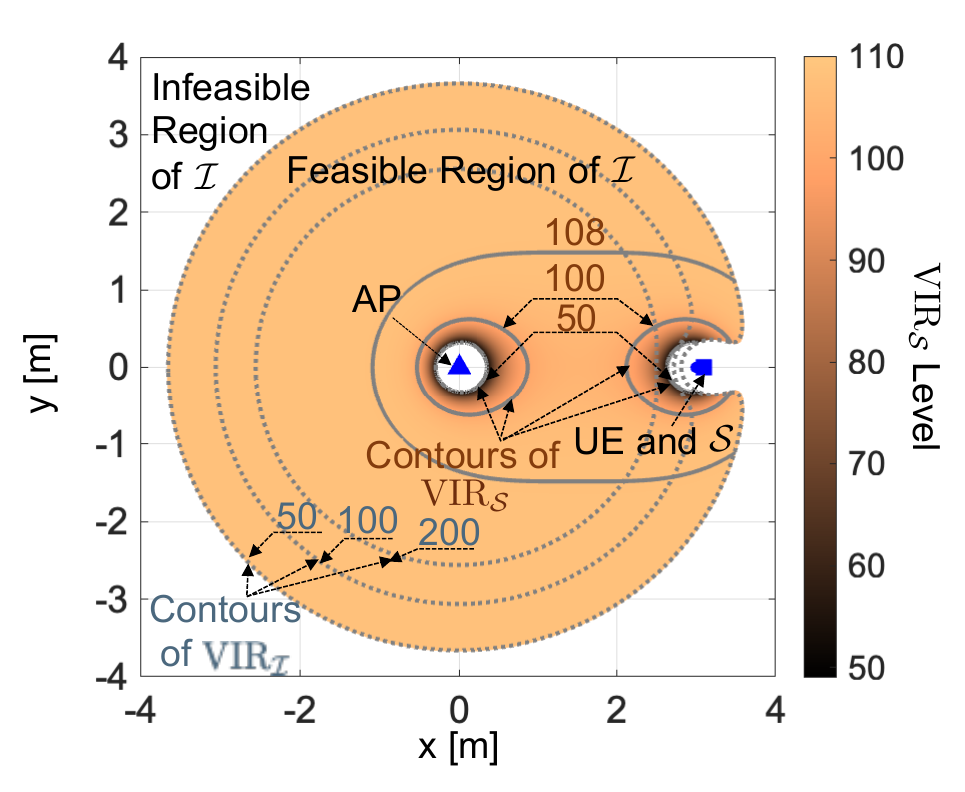}
    \caption{Feasible region of \gls{int} and the contours of VIR levels for both \gls{tar} and \gls{int}.} 
    % at different positions.}
    %\vspace{-1.5ex}
    \label{fig: vsir map}
    \vspace{-.5ex}
    % [Jingzhi Hu, 2023/3/8 21:13, (NOTE)]: 上面的图形实际上就是双基站雷达的Ovals of Cassini. 也是和Zhang Daqing他们文章里面的图是具有一致的特点的。
\end{figure}

To deliver more visual insights, we illustrate the feasible region of \gls{int} for $\gls{threshold}=50$ in Figure~\ref{fig: vsir map}, given $v_{\gls{tar}}$, $v_{\gls{int}}$, $\gamma$, $b$, $\tilde{G}_{A,\gls{tar}, E}$, and $\tilde{G}_{A,\gls{int},E}$ being normalized to $1$ and $\alpha=4$.
It can be observed (e.g., by the small infeasible circular regions around \gls{tar} and the AP) that the separation distance between \gls{tar} and \gls{int} can be potentially short without resulting in poor VIR for \gls{tar}, given \gls{int} is not close to the AP, too.
Moreover, as \gls{int} is also a potential subject of the Wi-Fi sensing system, it needs to be treated symmetrically to \gls{tar}, i.e., \gls{int} is also interfered by \gls{tar}, so $\mathrm{VIR}_{\gls{int}}$, similarly characterized as~$\mathrm{VIR}_{\gls{tar}}$ in~Eqn.~\eqref{equ: vsir}, should also be dominating at \gls{int}'s UE, meaning $\mathrm{VIR}_{\gls{int}}<\gls{threshold}$ when $d_{A,\gls{int}}$ being large is infeasible too. 
Therefore, besides the small regions mentioned earlier, a bigger one around the whole system indicates the whole boundary of \gls{int}'s feasible region.

Although these results are obtained numerically and serve for indicative purpose only, the resulting feasible region for a single interferer \gls{int} establishes the physical guarantee that the channel variation due to \gls{tar} can be well separated from that due to \gls{int}, and vice versa.
It can also be observed from Figure~\ref{fig: vsir map} that the contours of $\mathrm{VIR}_{\gls{tar}}$ constitute a set of \emph{Cassini oval}s around the AP and \gls{tar}, which may appear to be similar to the experimental results in \cite{Wang2022Placement}.
Nevertheless, our VIR is derived from the perspective of channel variation rather than from the SNR metric adopted in \cite{Wang2022Placement}, which evaluates the ratio between the power of signals reflected by the subject and the noises. 
We strongly believe our channel variation analysis is far more relevant to sensing applications, as the physical information of a subject is represented by the channel variation rather than the signal strength.

\subsection{Insights into Multi-person Scenarios}
\label{s2ec: imp to mul per scene}
Now we are ready to extend the analysis in Section~\ref{s2ec: near-field domination} to multi-person scenarios, where $N$ subjects~($N\geq 3$) are using the Wi-Fi sensing system and each of them stays in the near-field of its UE.
Though we could extend Eqn.~\eqref{equ: vsir} to multi-person case by putting $N$-fold interference in the denominator, the resulting characterization would be too general to shed any immediate lights on sensing performance, because the distribution patterns of these subjects have infinite possibilities.
To this end, we consider two symmetric distribution cases, aiming to address two important questions separately: i) how many subjects can a Wi-Fi sensing system support? and ii) how close can adjacent subjects be?
To simplify the presentation, we \rev{no longer} distinguish between the positions of the subject and its UE.

\begin{figure}[t]
    \setlength\abovecaptionskip{0pt}
     \setlength\subfigcapskip{-2pt}
    \centering
    \subfigure[Radial symmetry case.]{\label{fig: two special cases i}
	    \includegraphics[width=.35\linewidth]{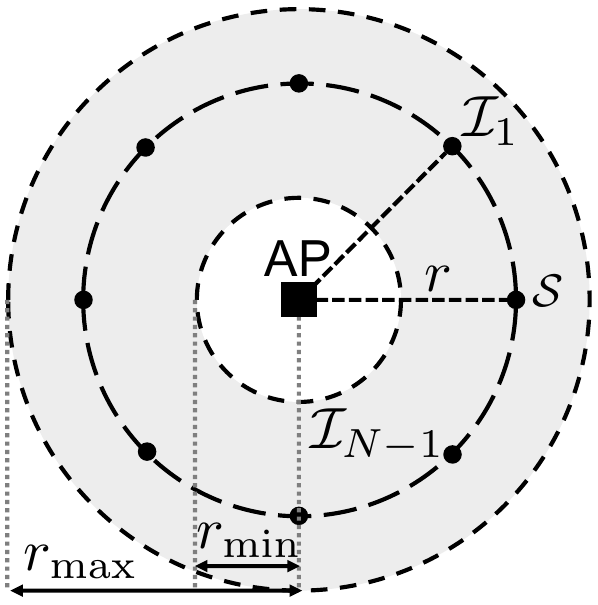}
	    \label{sfig:radialcase}
    }
    \hspace{5ex}
	\subfigure[Mirror symmetry case.]{\label{fig: two special cases ii}
		\includegraphics[width=.35\linewidth]{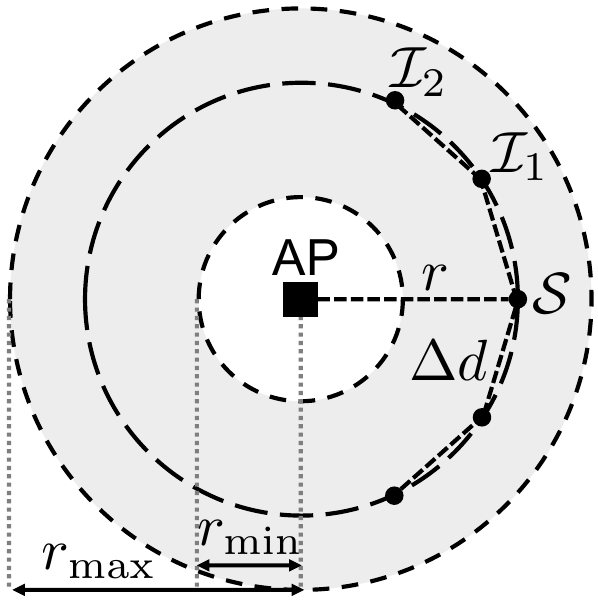}
		\label{sfig:mirrocase}
	}  \\[-0.3em]
    \subfigure[$N_{\max}$ vs $r$.]{\label{fig: N max d min 1}
    	\includegraphics[width=0.475\linewidth]{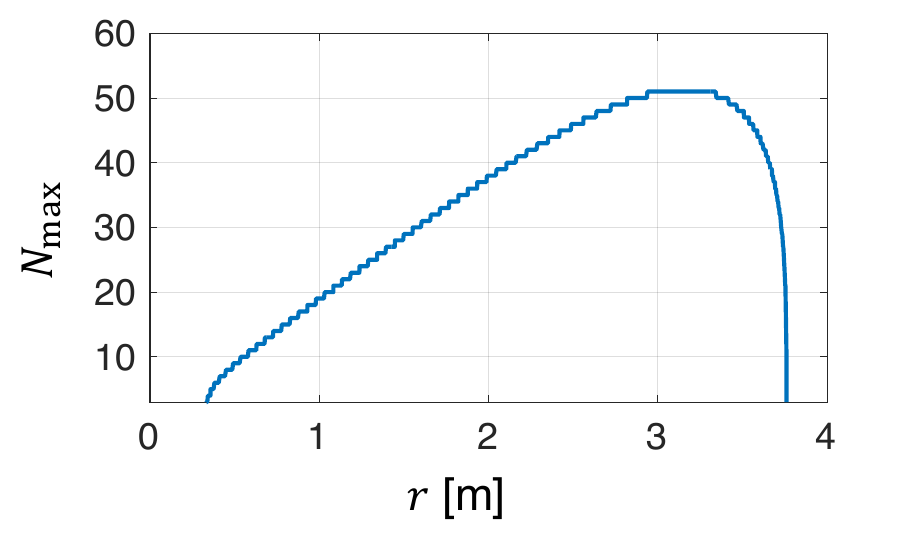}
    	\label{sfig:nmax}
         \vspace{-2ex}
    }
    \hfill
    \subfigure[$d_{\min}$ vs $r$.]{\label{fig: N max d min 2}
    	\includegraphics[width=0.45\linewidth]{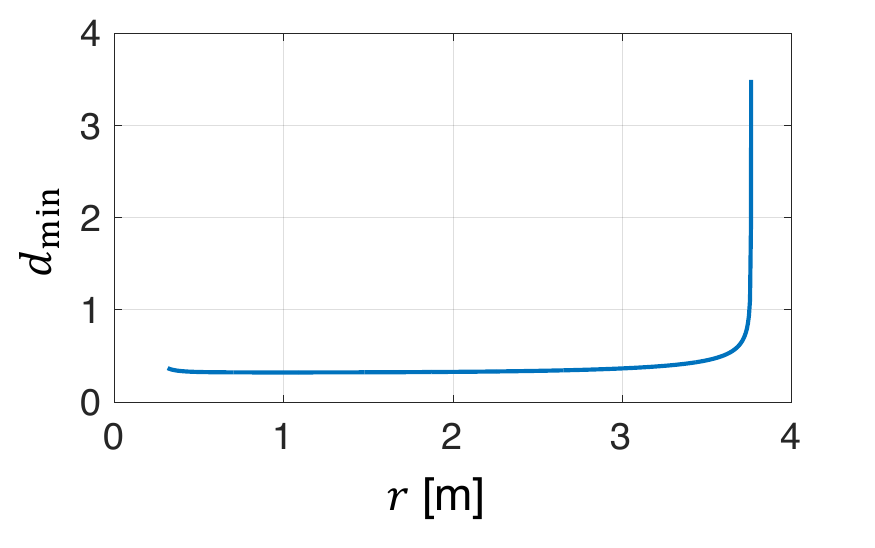}
    	\label{sfig:dmin}
      \vspace{-2ex}
    }
    \caption{Two symmetric cases considered for multi-person sensing scenarios. (a)\&(c) $N$ subjects uniformly spaced and (b)\&(d) $2K+1$ subjects closely located.}
    \label{fig: two special cases}
    \vspace{-2ex}
\end{figure}

To answer question i), we analyze the case where $N$ subjects stay at distance $r$ from the AP and are uniformly spaced, as shown in Figure~\ref{sfig:radialcase}.
Due to the radial symmetry of their positions, we can focus on analyzing any of them, which is again named \gls{tar}.
Extending Eqn.~\eqref{equ: vsir}, we have:
\begin{equation}
\label{equ: case a vsir}
\mathrm{VIR}_{\gls{tar}}^{(\text{i})} = \frac{\tilde{G}(\Delta r)^{-\alpha}}{\eta \lambda^2 + b r^\alpha +(2r)^{-\alpha}\tilde{G}\cdot\sum_{j=1}^{N-1} \sin^{-\alpha}(j\cdot \pi/N)},
\end{equation}
where $\tilde{G}$ represents the identical values of $\tilde{G}_{A,\gls{tar},E}$ and $\tilde{G}_{A,\gls{int_j},E}$ ($\forall j\in\gls{setInterferer}$), and $\Delta r$ denotes the short distance from \gls{tar} to its UE.
Generally, the series summation in~Eqn.~\eqref{equ: case a vsir}, $\sum_{j=1}^{N-1} \sin^{-\alpha}(j\cdot \pi/N)$, has no closed-form expression w.r.t. $N$.
Fortunately, given $N\in[3,60]$ and $\alpha\in [2,4]$, the series summation can be numerically fitted by a function in the form of $p_1N^{p_2} + p_3$ with $\text{R-square}\approx 1$, where parameters $p_1$, $p_2$, and $p_3$ are dependent on $\alpha$:
$p_1=0.0230$, $p_2=3.99$, $p_3=38.0$ for $\alpha=4$.
Now given \gls{threshold}, the \textbf{upper bound} on the number of subjects that can be accommodated becomes:
\begin{equation}
\label{equ: N_max}
\!\!\!\!N_{\max}\!=\!\left\lfloor\left({\frac{(2r)^{\alpha}}{p_1}\cdot\frac{\tilde{G}(\Delta r)^{-\alpha} - \eta\lambda^2\gls{threshold} - br^{\alpha}\gls{threshold}}{\tilde{G}\gls{threshold}} - \frac{p_3}{p_1}}\right)^{1/p_2}\right\rfloor.\!\!\!
\end{equation}
Moreover, based on~(\ref{equ: N_max}), we can also derive the minimum and maximum distances (resp. $r_{\min}$ and $r_{\max}$) between \gls{tar} and the AP for the considered case to be feasible by solving the inequality $N_{\max} \geq 3$ in the field of real number.
As shown in~Figure~\ref{sfig:nmax}, $N_{\max}$ first increases and then decreases in $r$; it reaches its maximum $51$ when $r\in[2.94,3.35]$~m.

To answer question ii), we consider the case as shown in Figure~\ref{sfig:mirrocase}, where $N=2K+1, K=1,2,3,...$ .
Denote the distance and angle between each pair of neighboring subjects by $\Delta d$ and $\phi$, we have $\phi = 2\cdot \asin(\Delta d/(2r))$.
It is clear that the middle subject \gls{tar} suffers from the worst interference (among all subjects) when $2\pi-(2K+1)\phi>0$, equivalent to $\Delta d<2r\sin({\pi}/({2K+1}))$.
Therefore, given $\Delta d<2r\sin({\pi}/({2K+1}))$, the condition for $\mathrm{VIR}_{\gls{tar}} > \gls{threshold}$ determines the low bound of $\Delta d$. Using Eqn.~\eqref{equ: vsir} again, we have:
\begin{equation}
\mathrm{VIR}_{\gls{tar}}^{(\text{ii})} = \frac{\tilde{G}\Delta r^{-\alpha}}{\eta\lambda^2 + br^{\alpha} + 2\tilde{G}(2r)^{-\alpha}\sum_{j=1}^K \sin^{-\alpha}(j\phi/2)}.
\end{equation}
Again, the series summation $\sum_{j=1}^K \sin^{-\alpha}(j\phi/2)$ can be fitted by function $q_1(\sin(\phi/2))^{q_2}+q_3$ given $\alpha\in[2,4]$, $K\in[1,10]$, and $\phi\in[\pi/180, \pi/(2K+1)]$ with $\text{R-square} \approx 1$, where parameters $q_1$, $q_2$, and $q_3$ depend on $K$ and $\alpha$: $q_1=1.06$, $q_2=-4$, and $q_3=6.57$ for $\alpha=4$ and $K=2$. 
Consequently, we obtain the \textbf{lower bound} of $\Delta d$ as follows:
\begin{equation}
\!\!\!\Delta d_{\min} = 2r\left({\frac{(2r)^{\alpha}}{q_1}\cdot\frac{\tilde{G}(\Delta r)^{-\alpha} - \eta\lambda^2\gls{threshold} - br^{\alpha}\gls{threshold}}{2\tilde{G}\gls{threshold}} \!-\! \frac{q_3}{q_1}}\right)^{\frac{1}{q_2}}.\!\!
\label{eq:deltadm}
\end{equation}
By solving $\Delta d_{\min}\leq 2r\sin(\pi/(2K+1))$ in the field of real number, we can obtain the boundary for the distance between the AP and the subjects, i.e., $r_{\min}$ and $r_{\max}$.
Figure~\ref{sfig:dmin} shows that $\Delta d_{\min}$ remains around $0.34$~m for $r\in[0.32,3.30]$~m but increases steeply to $3.49$~m for $r\in[3.30,3.76]$~m.

\subsection{Proof-of-Concept Pre-Experiments} 
\label{ssec:imp-mul}

\begin{figure}[t]
    %\vspace{-2ex}
    \setlength\abovecaptionskip{3pt}
     \setlength\subfigcapskip{-2pt}
    \centering
    \subfigure[Setting one.]{
    	\label{fig: 4 people sensing scene}
	    \includegraphics[width=0.39\linewidth]{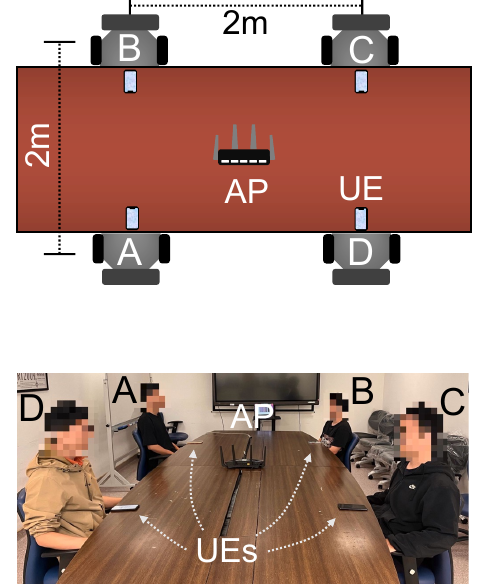}
	    \vspace{-2ex}
    }
    \hspace{-1ex}
    \subfigure[Multi-person respiration sensing.]{
    	\label{fig: 4 people sensing result}
    	\includegraphics[width=0.545\linewidth]{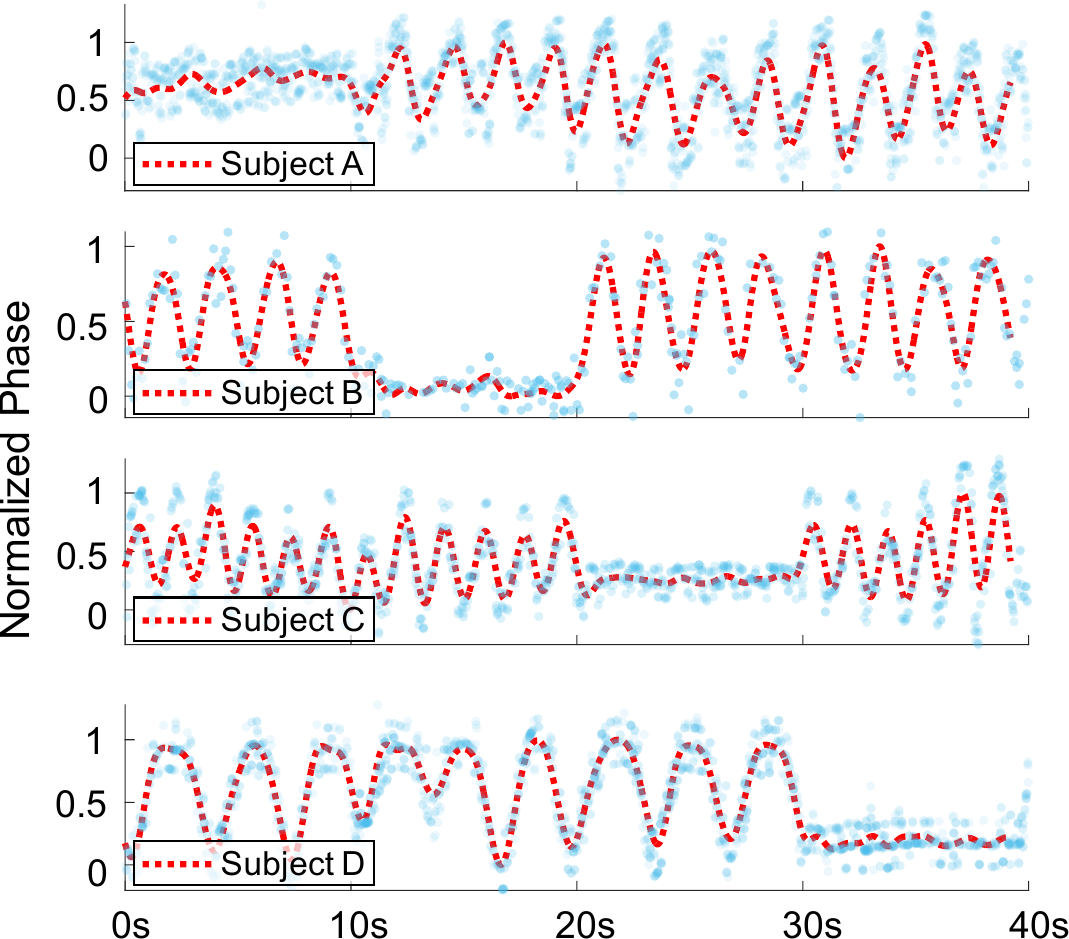}
    	\vspace{-2ex}
    }
    \\[-.2em]
    \subfigure[Sitting and breathing.]{
		\label{fig: two people breath}
		\includegraphics[width=0.465\linewidth]{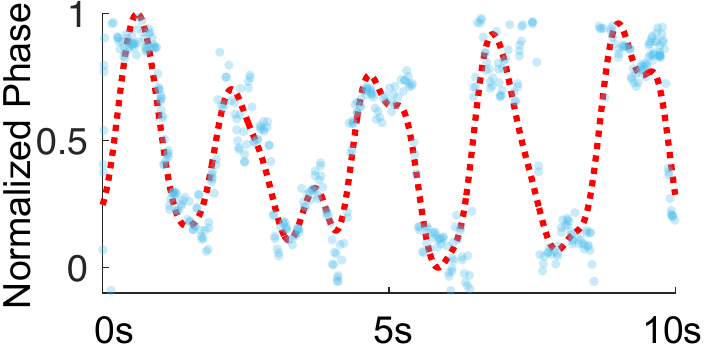}
		\vspace{-2ex}
	}
	\hfill
	\subfigure[Standing up and sitting down.]{
		\label{fig: two people sit}
		\includegraphics[width=0.465\linewidth]{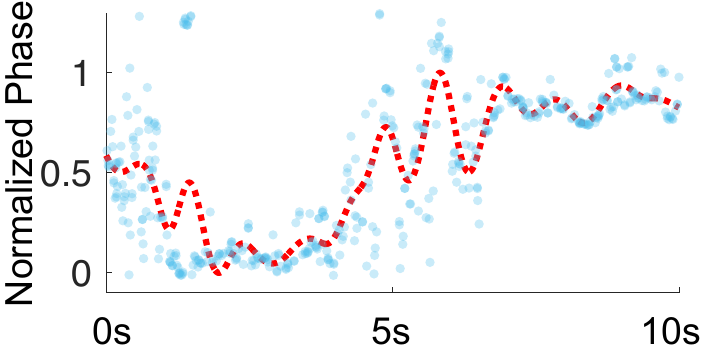}
		\vspace{-2ex}
	}
    \subfigure[Setting two.]{
		\label{fig: three people sit scene}
		\includegraphics[width=0.3\linewidth]{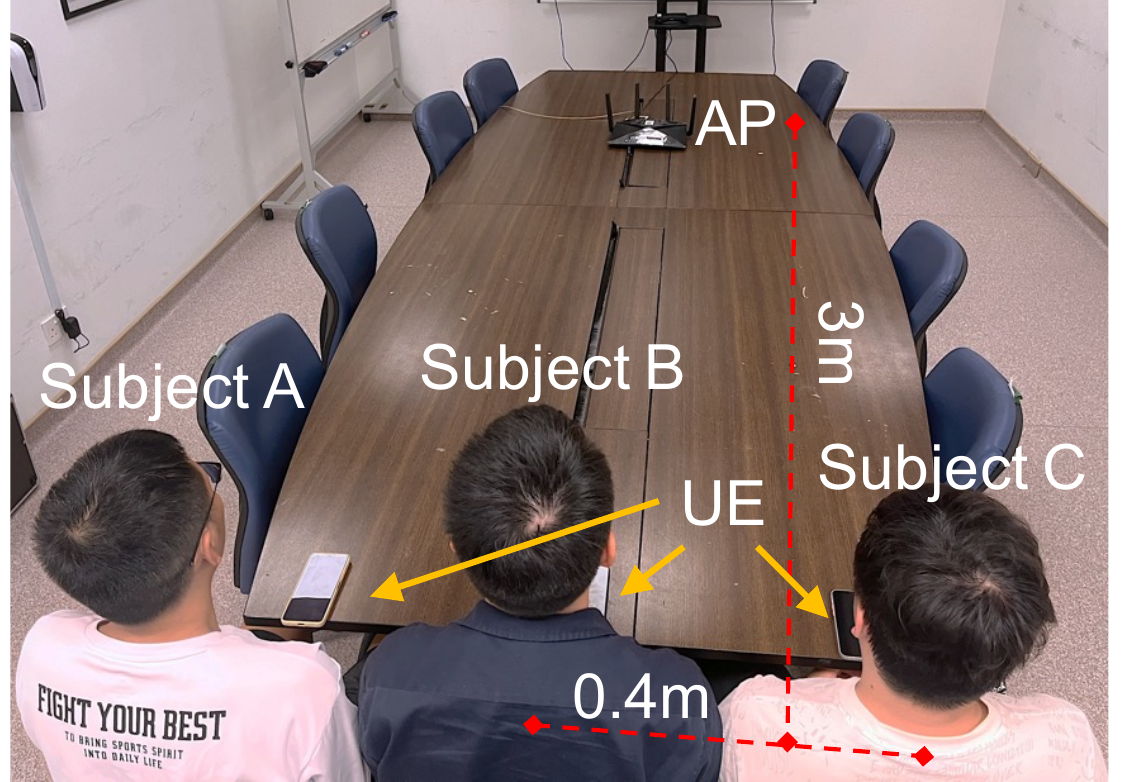}
		\vspace{-2ex}
	}
    \hspace{-.8ex}
    \subfigure[Respiration of the middle subject.]{
		\label{fig: three people sit}
		\includegraphics[width=0.65\linewidth]{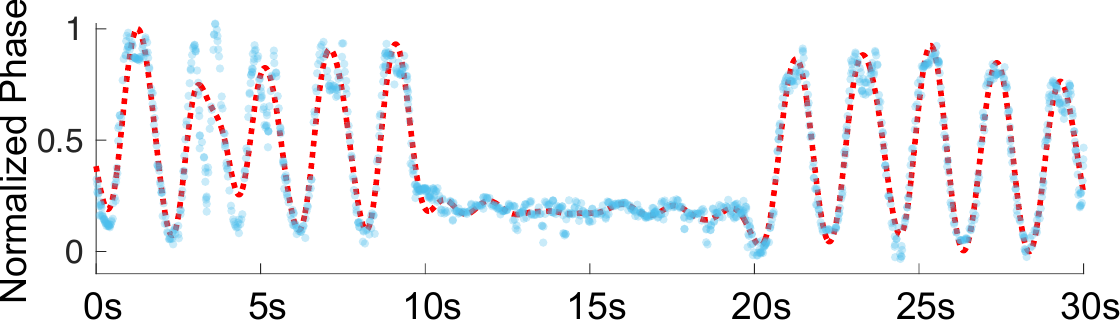}
		\vspace{-2ex}
	}
    \caption{Preliminary experiments. (a) Setting for the first two experiments. (b) Multi-person respiration sensing in action. (c) and (d) Multi-person asymmetric sensing with both respiration and activity. (e) Setting for the 3rd experiment. (f) Respiration (held or not) from the middle subject, i.e., Subject B in (e).}
    \label{fig: 4 people sensing}
    \vspace{-1.5ex}
\end{figure}

We first conduct two preliminary experiments to briefly validate the theoretical analysis in Eqn.~(\ref{equ: N_max}). 
We deploy a Wi-Fi~5 AP in the middle of a table, serving $4$ users spaced $2$~\!m apart around the table, as depicted in Figure~\ref{fig: 4 people sensing scene}. 
Each \emph{user} \rev{in this paper} involves a UE and a subject placed 15~\!cm apart from each other. 
We leverage the iPerf3 tool~\cite{tirumala1999iperf} to emulate the $4$ UEs and the AP constantly exchanging data frames with each other for 40 seconds, during which the four subjects take turns holding their breath. 
Their irregular CSI sequences (blue points) are obtained by the uplink traffic from the UEs to the AP, and the outcomes after the processing of \name are shown as red (dotted) curves in Figure~\ref{fig: 4 people sensing}.

Our results in Figure~\ref{fig: 4 people sensing result} show that the respirations of the $4$ subjects never interfere with each other, indicating an effective multi-person sensing. 
Though the theoretical results in Figure~\ref{sfig:nmax} suggest that up to 25 users can be supported with $r = 1.41$~\!m, our preliminary experiments are conducted in a more conservative manner, as the theoretical results are obtained under ideal conditions without user asymmetry. 
This setting also leaves room for us to validate asymmetric sensing, where we let three subjects breathe normally while the Subject~D performs the activity of standing up and sitting down. 
For brevity, we only show the sensing results for one of the breathing subjects and Subject~D in Figures~\ref{fig: two people breath} and \ref{fig: two people sit}, respectively; these results clearly demonstrate that multi-person sensing can still be successfully achieved even under asymmetry scenarios. 
In our case studies later, more users will be involved to better confirm the effectiveness of near-field domination for \name.

We conduct another experiment to validate the case of close proximity between users in Eqn.~\eqref{eq:deltadm}. 
We let three subjects sit side by side with $40$~\!cm distance between the centers of gravity of the neighboring ones, as shown in Figure~\ref{fig: three people sit scene}. 
Our results in Figure~\ref{fig: three people sit} focus on the \rev{centered} subject (the most interfered one), whose respiratory (held or not) can be clearly sensed regardless of the behaviors of others. 
These results evidently validate the effectiveness of feasibility of near-field domination on Wi-Fi multi-person sensing even under close proximity between neighboring subjects.

%% file: 3_system_design.tex
Given the physical guarantees on multi-person separation through near-field domination, we hereby introduce the detailed components and sensing strategies of \name.
In hardware, \name is comprised of a commodity AP and multiple UEs owned by subjects demanding sensing service\rev{s} from the system; all Wi-Fi devices follow the prevalent Wi-Fi standard and their CSIs can be readily obtained from the received frames~\cite{Nexmon-WiNTECH19,PicoScenes-IoIJ21}.
In the following, we first introduce three sensing strategies adopted by \name, along with their practical issues. We then specifically investigate two critical issues faced by these strategies.

\subsection{Three Sensing Strategies for \name} \label{ssec:three}
Since CSIs are carried by Wi-Fi frames, \name has three sensing strategies based on the traffic direction and how CSIs are carried. In particular\rev{, they are:}
\begin{itemize}[leftmargin=*]
\item \textbf{UL-CSI Sensing}: The uplink~(UL) traffic and the CSI, obtained from the long training sequence (LTS) carried in the preamble of a frame, are adopted as sensing primitives.  
\item \textbf{DL-CSI Sensing}: The downlink~(DL) traffic and the carried CSI are adopted as sensing primitives.
\item \textbf{UL-BFI Sensing}: The \rev{UL} traffic and the carried BFI are adopted as sensing primitives.
\end{itemize}

Apparently, the entity to handle sensing (data processing) depends on the direction of data flow: both UL-CSI and UL-BFI require sensing to take place on the AP side, while DL-CSI demands the UE to handle sensing.
Here, BFI is actually a compressed version of the DL-CSI but carried by UL traffic for the AP to be aware of the DL channel conditions, so as to fine-tune its MIMO precoding; it only becomes available since IEEE 802.11ac standard~\cite{wifi80211}. 
As BFI is transmitted with \textit{action frame}s (part of UL traffic) in plain form, sensing can also take place at Wi-Fi devices capable of overhearing the traffic; the incurred privacy issue will be further discussed in one of our companion works.

\vspace{-.5em}
\subsubsection{User Registration}\label{ssec: user registration}

When a \textit{user} (a subject with its UE) demands access to \name, it should first announce its presence along with the sensing application it requires.
This user registration process is necessary for three reasons: 
i) it lets \name be aware of \rev{the number of users} and their respective motion types, so as to coordinate users more accurately (e.g., reject users if system capacity is reached), 
ii) it prepares \name with prior information to fine-tune its processing pipelines, such as selecting different filters according to \rev{motion intensity} or involving algorithmic modules to handle excessive interference if the average \rev{motion intensity} is too high,
and iii) it preserves privacy for normal Wi-Fi users with no \rev{intention} to access \name.
In the following, we \rev{will} focus only on the registered users, leaving detailed registration procedure in our extended report.

\vspace{-.5em}
\subsubsection{Practical Issues}

\begin{figure}[b]
    \vspace{-1em}
    \setlength\abovecaptionskip{0pt}
     \setlength\subfigcapskip{-2pt}
    \centering
	\subfigure[One User.]{
		\label{fig: frame arrival rate a}
		\includegraphics[width=0.465\linewidth]{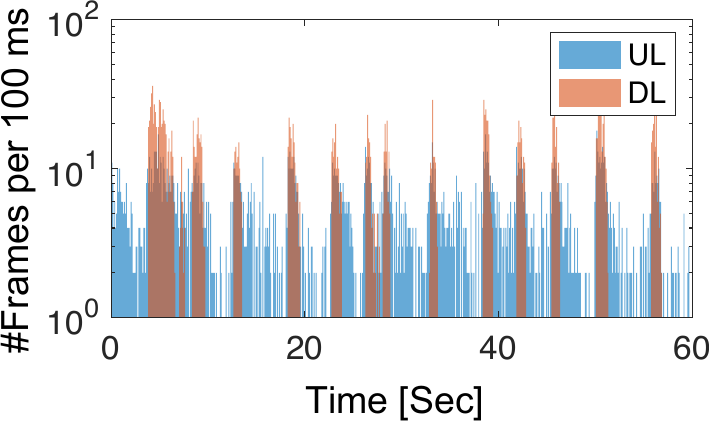}
		\vspace{-2ex}
	}
	\hfill
	\subfigure[Two Users.]{
		\label{fig: frame arrival rate b}
		\includegraphics[width=0.47\linewidth]{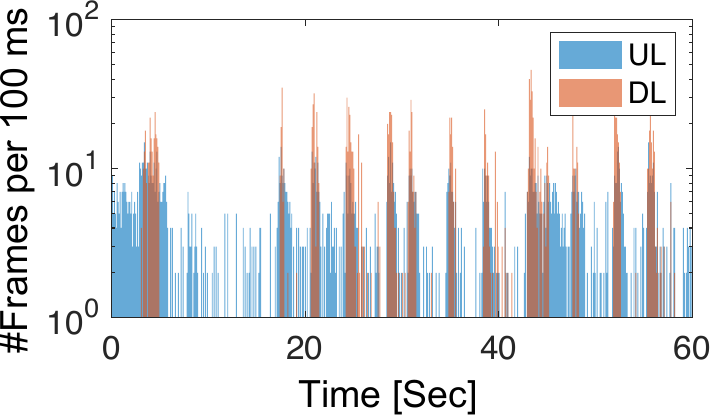}
		\vspace{-2ex}
	}
    \caption{Frame arrival rates in terms of number of frames per 100~\!ms versus the observation time when (a)~one or (b)~two users stream 1080p videos.}
    \label{fig: frame arrival rate}
\end{figure}

In stark contrast to existing Wi-Fi sensing systems working with artificially generated continuous \rev{traffic} and mostly with only a single link, the most prominent challenge for \name is to handle the bursty and intermittent traffic in practice.
Specifically, due to the multi-user communication infrastructure and the contention-based medium access mechanism on which \name is based, both UL and DL \rev{traffic} exhibit bursty and intermittent characteristics, leading to CSI time series \rev{being} sparse with many discontinuous parts.
To illustrate this, we depict the frame arrival rates for both UL and DL when one or two users watch 1080p videos using their respective UEs in Figure~\ref{fig: frame arrival rate}. 
Both UL and DL traffics already exhibit bursty and intermittent patterns for one user, caused by upper-layer protocols' data caching and rate control~\cite{Wu2001Streaming}; these are exacerbated by channel contentions even with only one additional user, as shown in Figure~\ref{fig: frame arrival rate b}.
Moreover, the BFI is contained only in a small portion of UL traffic, and its sample rate is about $1/10$ of DL frames, peaking at roughly $10$~frames \rev{per} second, which means the UL-BFI sensing strategy faces the severest data sparsity.
Consequently, \name needs to be capable of recovering continuous channel variation from a CSI time series with sparse samples.

\vspace{-.5em}
\subsection{Sparse Recovery Algorithm} 
\label{ssec:sparse} 
We propose an SRA for \name to recover continuous channel variation from the intermittently sparse samples due to realistic traffic.
The SRA is comprised of two components: a \emph{data transformation pipeline} to pre-process a sparse CSI sequence sampled under realistic traffic, and a \emph{self-supervised data recovering network} to recover the densely sampled CSI sequence.
\frev{The core novelty of SRA lies in eliminating the extensive label collection, by interpreting the correlations between sparse and non-sparse data slices.}
Without loss of generality (of three sensing strategies), we denote the input data to SRA by a 1D time series $\{{x}_t\}$ drawn from the \rev{CSI} of specific antenna pair and subcarrier, where $t$ is the sampling time and \rev{${x_t}\in\mathbb{R}$} is the phase of a corresponding CSI sample.
SRA outputs $\{\bm y_t\}$ as an evenly and densely sampled multi-channel sequence where $\bm y_t\in\mathbb R^{\gls{numFreqCp}}$ represents the \gls{numFreqCp} frequency components of the CSI \rev{for} time $t$.
This output can be used directly as the sensing result, or be further processed to recognize the activity or gesture of the subject.
We elaborate on the two SRA components in the following, .

\subsubsection{Data Transformation Pipeline}

\begin{figure}[t]
    %\vspace{-.5ex}
    \setlength\abovecaptionskip{3pt}
    \centering
    \includegraphics[width=0.8\linewidth]{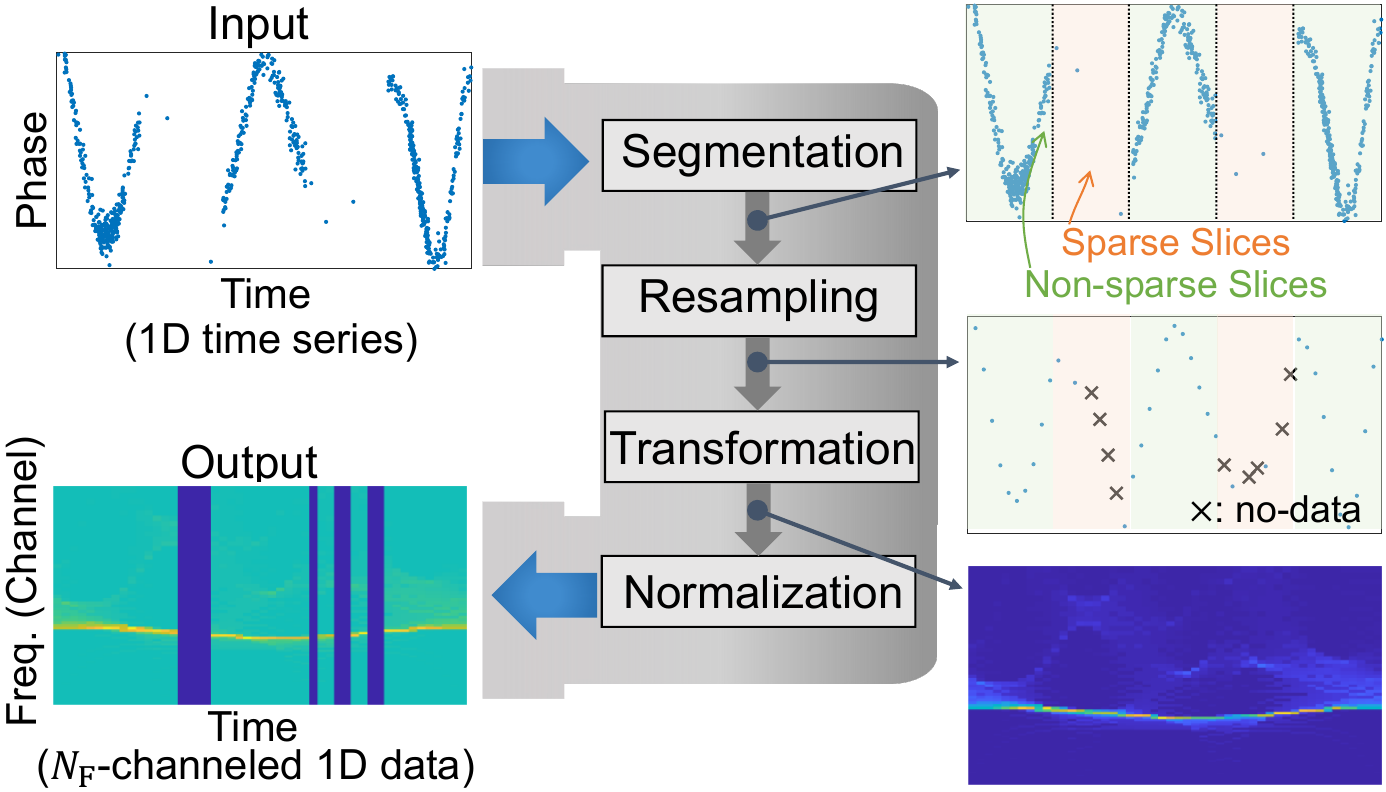}
    \caption{\rev{Data transformation  pipeline of \name.}}
    \label{fig: pipeline}
    \vspace{-3ex}
\end{figure}

This four-step pipeline transforms $\{{x}_t\}$ into an evenly resampled output sequence $\{\hat{\bm{x}}_n \in [-1,1]^{\gls{numFreqCp}}\}_{1\leq n \leq{\gls{numSample}}}$ with length $\gls{numSample}$ as the total number of resampled time instants, which is shown in Figure~\ref{fig: pipeline}.
\vspace{-.5em}
\paragraph{Segmentation}
We first segment the time series into two types of slices, i.e., \emph{sparse slices}, where the samples are sparse in time, and \emph{non-sparse slices}, for them to be treated differently.
This segmentation is done by using a sliding window of length $\gls{nspThreshold_t}$ in time to check whether it contains a sufficient number of samples. 
In particular, time \rev{slices} with more than $\gls{nspThreshold_N}$ samples are marked as \emph{non-sparse}; otherwise as \emph{sparse}.
Here \gls{nspThreshold_t} and \gls{nspThreshold_N} are parameters specified by sensing applications and are empirically set in Section~\ref{sec:implementation}.

\vspace{-.5em}
\paragraph{Resampling}
The time series is resampled so that the samples can be evenly spaced in time, facilitating further denoising and sparse recovery.
Within each non-sparse slice, the outliers are removed, and an interpolation is performed to meet a resampling frequency \gls{freq_resample}.
Each sparse slice is resampled with frequency \gls{freq_resample}, with samples moved to their nearest resampled time instants; those time instants without data are tagged as ``\texttt{no-data}'' \rev{and filled linearly}. 
The result is denoised through a low-pass filter with cut-off frequency \gls{cutoffFreq}.
Both \gls{freq_resample} and \gls{cutoffFreq} are empirically specified in Section~\ref{sec:implementation}.

\vspace{-.5em}
\paragraph{Transformation} % Or decomposition
This step transforms the resampled time series into its \emph{spectrogram} \rev{$\{\tilde{\bm{x}}_n \!\in \!\mathbb{R}^{\gls{numFreqCp}}\}_{1\leq n \leq{\gls{numSample}}}$ with $N_{\mathrm{F}}$} frequency components.
The reason behind this is that motions of subjects generally lead to channel variations whose patterns are environment- and subject-specific and hardly \rev{recognizable in the time domain}.
By transforming it into a spectrogram, the impact of subject's motion on \rev{the} channel becomes more apparent, facilitating effective sparse recovery.

\vspace{-.5em}
\paragraph{Normalization}
Mapping each \rev{$\tilde{\bm{x}}_n$} into $\hat{\bm{x}}_n \in [0,1]^{\gls{numFreqCp}}$ via min-max normalization allows for focusing on the relative variation pattern \rev{while} eliminating the magnitude difference of channel variation potentially caused by subject positions.
Besides, the \rev{frequency components} of time instants with \texttt{no-data} tags are assigned value $-1$, making them distinct from those with data and clearly indicating the data sparsity.

\subsubsection{TCN-based Sparse Slice Recovering}
Rather than employing a heavy neural network like U-Net for audio inpainting applications~\cite{kegler2019deep}, we adopt a temporal convolutional network~(TCN) based autoencoder~(AE) to achieve sparse recovery\rev{, involving} \rev{fewer} parameters \rev{to} make it efficient to train and deploy in resource-limited devices as UEs and APs.
TCNs are superior to other types of neural networks (e.g., LSTMs) as they exploit convolutional layers with dilated kernels to capture ultra long-range dependencies in samples while maintaining a manageable number of parameters~\cite{bai2018empirical}.

\begin{figure}[b]
    \vspace{-1ex}
    \setlength\abovecaptionskip{6pt}
    \centering
    \includegraphics[width=1\linewidth]{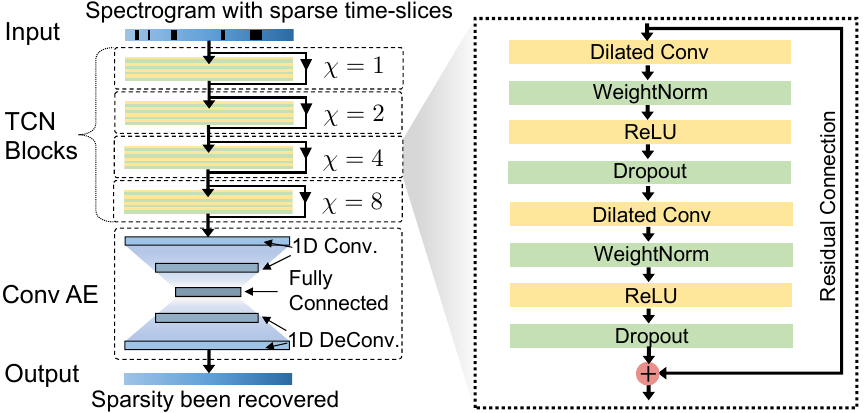}
    \caption{\rev{The structure of TCN-AE in \name.}}
    \label{fig: tcn}
    \vspace{-.5ex}
\end{figure}

\vspace{-.5em}
\paragraph{Network Structure}
As shown in Figure~\ref{fig: tcn}, the \rev{designed} TCN-based AE consists of an input layer, $4$ TCN blocks, a 1D convolutional AE module, and an output layer.
The core of the network is the TCN blocks, whose components are featured by the dilated convolutional layers and a residual connection.
In particular, taking the first dilated convolution layer as an example, it takes $\{\hat{\bm x}_n\}_{1\leq n\leq \gls{numSample}}$ as input and apply\rev{s} dilated convolution to it with \gls{numChannel} 1D-kernels to obtain a \gls{numChannel}-channeled output $\{\bm z_n\}_{1\leq n\leq \gls{numSample}}$ for the next layer.
For the $k$-th channel of the output~($k=1,...,\gls{numChannel}$), given the 1D-kernel $\bm F_{k} = (\bm f_{k,1},...,\bm f_{k,L})$ with $\bm f_{k,l}\in \mathbb R^{\gls{numFreqCp}}$~($l=1,...,L$) and $L$ being the kernel size, the dilated convolution can be expressed as:
\begin{equation}
\label{equ: tcn dilated conv}
z_{k,n} = \textstyle{\sum_{i=0}^{L-1}} \bm f_{k,i+1}^{\top} \hat{\bm x}_{n-\chi\cdot i},~\forall n\in \{1,...,\gls{numSample}\},
\end{equation}
where $\hat{\bm x}_n$ is zero-padded for $n<1$, $(\cdot)^\top$ is the transpose operator, and $\chi\in\mathbb Z^{+}$ denotes the \emph{dilation factor} used to expand the receptive field of \rev{the} output element.

According to~Eqn.~\eqref{equ: tcn dilated conv}, the operation with a small dilation factor (e.g., $\chi=1$) degenerates to a traditional convolution for extracting the features of local context around each element of the input. 
As $\chi$ increases, the mutual dependency of local features can be captured by the kernel, and each element of the output can represent the local features of input in a wider range.
Therefore, by utilizing a stack of dilated convolutional layers with exponentially increased $\chi$ as shown in Figure~\ref{fig: tcn}, the local features are gradually extracted and collected, enabling each node of the output layer to take into account well-represented local features for almost the entire spectrogram.
Finally, with the results of the TCN blocks, the AE can effectively predict and recover the missing data.
Representing the TCN-AE network as a ${\bm w}$-parameterized function $\mathcal F_{\bm w}$,
re-arranging $\{\hat{\bm x}_n\}_{1\leq n\leq \gls{numSample}}$ into matrix form $\hat{\bm X}\in\mathbb R^{\gls{numFreqCp}\times\gls{numSample}}$,
and denoting the output by $\tilde{\bm Y}\in \mathbb R^{\gls{numFreqCp}\times\gls{numSample}}$, the recovering process can be represented as $\mathcal F_{\bm w}: \hat{\bm X} \rightarrow \tilde{\bm Y}$.

\vspace{-.5em}
\paragraph{Self-supervised Training}

Generally, the pre-collected training dataset needs to contain data-label pairs: spectrogram with sparse slices as the data and corresponding ground truth with no sparsity as the label.
Unfortunately, collecting the ground truth \rev{directly} is almost impossible, as sparse slices are caused by the lack of frames (hence losing the carried ground truth samples) during certain periods.
\frev{To overcome this impossibility, we propose a self-supervised training method; it leverages only the non-sparse slices for training the TCN-AE, aiming to restore the hypothetical non-sparse data that facilitate various downstream sensing tasks.}
Specifically, we collect the spectrograms of non-sparse time slices to form the training label set, while \rev{obtaining} the corresponding input data by randomly assigning \texttt{no-data} tags to the elements for creating artificial data sparsity.
We note that this tag assignment needs to preserve the bursty and random patterns in the occurrence of \texttt{no-data} elements.

Moreover, to augment the training dataset, each expanded non-sparse spectrogram data in the label set are reused for \rev{multiple} times with random tag assignments.
Consequently, we obtain a completely labeled training dataset, denoted by $\mathcal D_{\mathrm{train}} = \{\mathcal T(\bm Y), \bm Y\}$, without resorting to the impossible ground truth collection process.
Here ${\bm Y}\in \mathbb R^{\gls{numFreqCp}\times\gls{numSample}}$ represents the spectrogram data of a non-sparse time slice, and $\mathcal{T}(\cdot)$ is the random tag assignment.
Finally, we can express the training process by the following optimization problem, to minimize the expected mean-squared error~(MSE) between the recovered spectrogram and its label:
\begin{equation}
\label{opt: training}
\min_{\bm w} ~\mathop{\mathbb E}\limits_{(\mathcal T({\bm Y}), {\bm Y})\in\mathcal D_{\mathrm{train}}}\|\mathcal F_{\bm w}(\hat{\bm X}) - \bm Y\|_2^2,~\text{s.t.}~\hat{\bm X} = \mathcal{T}(\bm Y). 
\end{equation}
\frev{The overheads of the training is minor because no online labeling process is needed, and thus MUSE-Fi can collect the dataset automatically and conduct the training offline without incurring any real-time overheads.}

\subsection{To Compress or Not to Compress?} 
\label{ssec:bfi_analysis}

In this section, we specifically study the effectiveness of BFI-enabled sensing.
Consider a conventional CSI matrix $\bm H\in \mathbb C^{\gls{numRx}\times\gls{numTx}}$ for a given subcarrier of a DL link with \gls{numTx} antennas for Tx at AP and \gls{numRx} antennas for Rx at UE. 
Instead of directly feedbacking $\bm H$ to the AP, the UE piggybacks a compressive form of $\bm H$ (i.e., BFI) onto the UL traffic, containing only the necessary information for Tx beamforming.
Consider a channel state represented by ${\bm H}_0$, the UE first obtains the Tx beamforming matrix $\bm V$ by conducting the singular value decomposition~(SVD) on ${\bm H}_0$, i.e., ${\bm H}_0 = \bm U \bm S \bm V^{*}$ where $\bm U\in \mathbb C^{\gls{numRx}\times\gls{numRx}}$ and $\bm V\in\mathbb C^{\gls{numTx}\times\gls{numTx}}$ are unitary matrices, $\bm S \in \mathbb R^{\gls{numRx}\times \gls{numTx}}$ is a rectangular diagonal matrix with non-negative real values on the diagonal, and $(\cdot)^{*}$ denotes the conjugate transpose.
The UE then compresses the channel state by converting $\bm V$ into BFI, which is represented by a series of real angles, and sends it to the AP. 
\begin{figure}[t]
    \setlength\abovecaptionskip{8pt}
    \centering
    \includegraphics[width=0.7\linewidth]{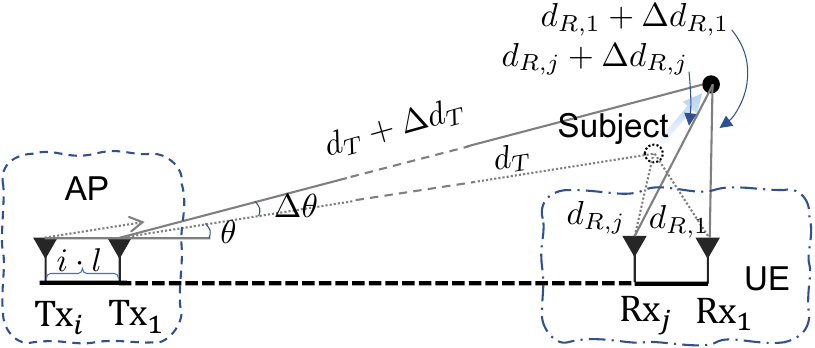}
    \caption{Channel variation due to subject motion: dashed and filled circles respectively represent the original subject position and that after motion.}
    \label{fig: svd analyze}
    \vspace{-1ex}
\end{figure}
With the BFI received at the AP, a \emph{reconstructed beamforming matrix} $\bm{\tilde{V}}$ is obtained, whose column vectors approximate those of $\bm V$ except for the column-wise phase-shifts that enforce the elements in the last row real-valued~\cite{survivalguide}.

Based on the premise above, the BFI-enabled sensing under \name's context can be analyzed, following the illustration in Figure~\ref{fig: svd analyze}, where a subject is in the near-field of the UE while far from the AP. 
After a displacement of the subject, the altered channel condition changes the CSI matrix to ${\bm H}_1$ and also affects the SVD result by
${\bm H}_1 = \bm Q_{\mathrm{rx}} {\bm H}_0 \bm Q_{\mathrm{tx}} = \bm{U}' \bm{S}'\bm{V}'^{*}$,
where 
$\bm Q_{\mathrm{rx}} =\diag(\rho_1 e^{-\iu\frac{2\pi}{\lambda}\Delta d_{R, 1}},..., \rho_{\gls{numRx}}e^{-\iu\frac{2\pi}{\lambda}\Delta d_{R, \gls{numRx}}})$
and 
$\bm Q_{\mathrm{tx}} = \diag( e^{-\iu \frac{2\pi}{\lambda} \Delta d_T}, ...,  e^{-\iu \frac{2\pi}{\lambda} \left[\Delta d_T - (\gls{numTx}-1) \ell \Delta\theta\sin(\theta)\right]})$ with $\ell$ being the distance between adjacent Tx antennas, and $\rho_{j}$ being the amplitude ratio between two channel gains respectively from the subject at new position and original position to the $j$-th Rx antenna. 
We can observe that $\bm{V}' = \bm Q_{\mathrm{tx}}^* \bm V$.
Thus, based on the relationship between $\bm{V}'$ and $\bm V$, the reconstructed beamforming matrix after motion becomes:
\begin{equation}
\label{equ: tilde V_}
\tilde{\bm V}' = \diag( e^{-\iu \frac{2\pi}{\lambda} (\gls{numTx}-1) l \Delta\theta\sin(\theta)}, e^{-\iu \frac{2\pi}{\lambda} (\gls{numTx}-2) l \Delta\theta\sin(\theta)}, ...,1) \tilde{\bm V}. \nonumber
\end{equation}

\begin{figure}[b] 
    \vspace{-3ex}
    \setlength\abovecaptionskip{0pt}
     \setlength\subfigcapskip{-2pt}
    \centering
	\subfigure[Respiration.]{
        \centering
		\label{fig:bfi_csi_re}
		\includegraphics[height=6em]{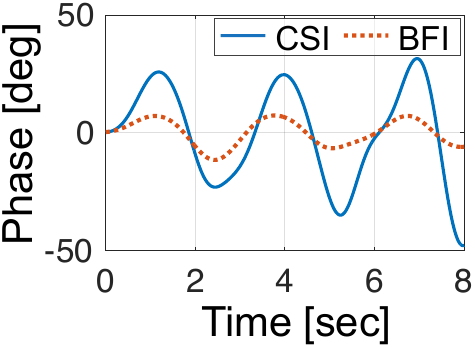}
		\vspace{-2ex}
	}
	\hspace{-.8em}
	\subfigure[\!\!Gesture (front-back).]{
        \centering
		\label{fig:bfi_csi_ge}
		\includegraphics[height=6em]{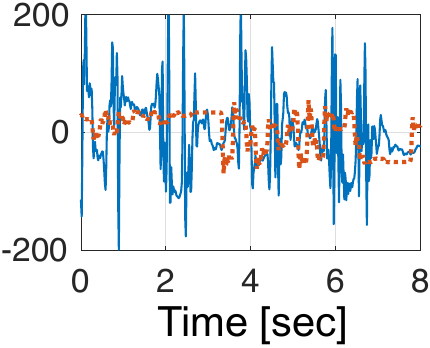}
		\vspace{-2ex}
	}
    \hspace{-.8em}
 	\subfigure[\!\!Activity (jumping).]{
        \centering
		\label{fig:bfi_csi_ac}
		\includegraphics[height=6em]{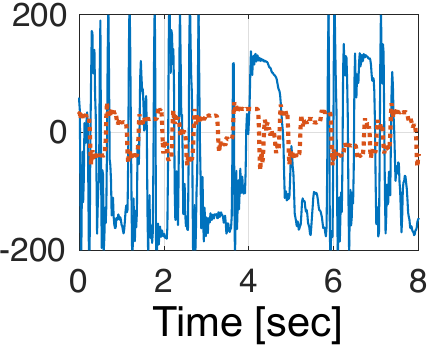}
		\vspace{-2ex}
	}
    \caption{CSI vs. BFI in the time domain.}
    \label{fig: bfi csi compare time}
    \vspace{-.5ex}
\end{figure}

Apparently, the BFI variation from $\tilde{\bm V}$ to $\tilde{\bm V}'$ depends only on the change of the relative direction from the subject to the AP, i.e., $\Delta \theta$, which does not concern the UE at all. 
As shown in Figure~\ref{fig: bfi csi compare time}, this compressive sensing brings \textbf{stability} to the sensing signal, yet at a cost of \textbf{reduced sensitivity} compared with CSI-based sensing, because BFI-sensing is almost insensitive to the relative motion between the subject and UE. 
For cases where the subject has rapid movements as shown in Figures~\ref{fig:bfi_csi_ge} and \ref{fig:bfi_csi_ac}, BFI sensing is more preferable as it produces results that are more stable, compared with CSI sensing that often causes drastic changes blended with noise and outliers. 
However, BFI sensing can be rather insensitive to micro-motions (albeit still viable), as demonstrated by Figure~\ref{fig:bfi_csi_re}. 
Therefore, whether to use CSI or compressed BFI depends on the specific application and the trade-offs between stability and sensitivity. Note that our analysis assumes that the subject is off the LoS path, which does not account for cases where, for example, the subject's hands are operating on a (smartphone) UE.

%% file: 4_implementation.tex
In this section, we first elaborate on \name's implementation, then we introduce the experiment setup. 

\subsection{Implementing \name} \label{ssec:implementation}

\name consists of an AP and multiple UEs owned by subjects seeking sensing services. The AP is a Netgear Nighthawk X10 router~\cite{netgear}, and the UEs include smartphones such as iPhone 13~\cite{iphone} and OnePlus 10T~\cite{oneplus}, as well as Acer TravelMate laptops~\cite{acer}. The Wi-Fi NICs adopted by \name employ 802.11b/g/n/ac for both UL-CSI and DL-CSI sensing, but utilize only 802.11ac for UL-BFI sensing (as BFI is available there only). The retrieval of CSIs is achieved via both Nexmon~\cite{Nexmon-WiNTECH19} and PicoScenes~\cite{PicoScenes-IoIJ21}, while Wireshark~\cite{orebaugh2006wireshark} is sufficient to obtain cleartext BFI information from \texttt{Action No-ACK} frames. The obtained CSI and BFI information is analyzed using Matlab. 

For training the SRA, after the sensing signals are passed through the pipeline, non-sparse slices in the spectrogram with a duration greater than $4$~\!s are picked for self-supervised training. To be specific, the non-sparse slices are used as ground truth (labels), and we then perform a random \texttt{no-data} tag assignment to them, thus generating sparse slices as the corresponding training inputs. We use 70\% of the slices for training TCN-AE and the remaining 30\% for testing. 
The parameters for sparse recovery are set as follows: $\Delta t = 0.1$~\!s, $N_{\mathrm{nsp}}= 2$, \rev{$L=5$}, $\gls{numFreqCp}=32$, $\gls{numChannel}=64$, $f_{\mathrm{rs}}=64$~\!Hz, and $f_{\mathrm{cut}}=1$~\!Hz for respiration monitoring $f_{\mathrm{cut}}=20$~\!Hz for other cases. 

\subsection{Experiment Setup} 
\label{ssec:setup}
\begin{figure}[b] 
    \vspace{-3ex}
    \setlength\abovecaptionskip{1pt}
     \setlength\subfigcapskip{-2pt}
        \hspace{-2em}
	\subfigure[Testing scene.]{
        \centering
		\label{fig: resp scene}
		\includegraphics[height=6.5em]{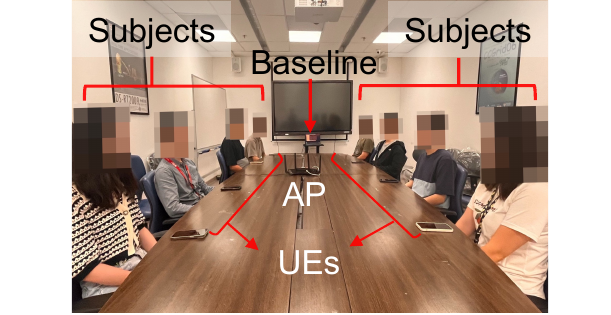}
		\vspace{-5ex}
	}
	\hspace{-.8em}
	\subfigure[Layout.]{	
        \centering
		\label{fig: resp layout}
		\includegraphics[height=6.5em]{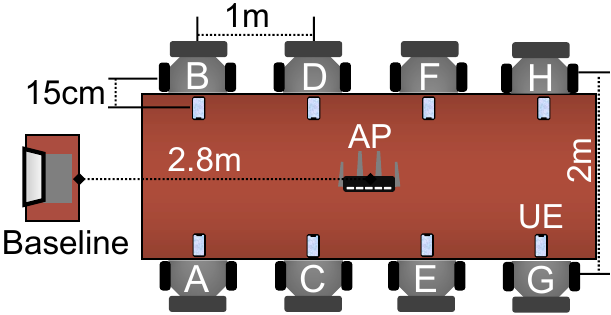}
		\vspace{-5ex}
	}
    \caption{Experiment scene (a) and layout of subject arrangement (b) for all three case studies. \frev{The 15cm UE-subject distance is only meant to indicate a near-field layout rather than be fixed to that given value.}}
    \label{fig: respiration scene layout}
    \vspace{-.5ex}
\end{figure}
We first conduct micro-benchmark studies with a real-time video conference application, then we perform three case studies for realistic sensing applications.
The setups for the case studies share three commonalities: i) each UE is placed in the near field of its associated subject, and it connects to the AP and continuously streams 1080p videos to emulate daily network usage, ii) all subjects perform specified activities simultaneously to test \name's ability in performing multi-person sensing, and iii) each experiment is conducted in a typical indoor meeting room with a different interior furniture arrangement.
We also compare \name with a \textit{non-near-field baseline} that employs another Wi-Fi device placed on the LoS path of the AP but not in the near-field of any subjects to collect CSI and BFI. 
Figure~\ref{fig: resp scene} illustrates our experiment setup for case studies, where the AP, subjects, UEs, and baseline device are all exhibited and annoted in Figure~\ref{fig: resp layout}.

\textit{Respiration Monitoring.}
We let 8 subjects breathe simultaneously, and use NeuLog chest belts~\cite{neulog} \rev{to obtain} the ground truth.
The total respiration recording period is 80-minute.  
During the \textit{Transformation} step of the SRA, the short-time Fourier transform~(STFT) is employed to focus on the low-frequency components of respiration
We employ a 3-layer convolutional neural network (CNN) to extract respiratory rate from the spectrogram.

\textit{Gesture Detection.} 
We let 8 subjects simultaneously perform six gestures, namely circle (CR), front-back (FB), slide (SL), star (ST), wave (WV), and zig-zag (ZZ). 
Each activity is performed 500 times, resulting in 24,000 CSI time series each containing 256 samples. We adopt the wavelet synchro-squeezed transform (WSST)~\cite{addison2017illustrated} in the \textit{Transformation} step, as it is highly effective in interpreting gesture signals that are non-stationary and contain complex frequency components. Besides, we employ the same classifier as in Widar3.0~\cite{Widar3-MobiSys19} to achieve gesture detection from the spectrogram.

\textit{Activity Recognition.}
We let 8 subjects simultaneously perform six daily observed human activities: bending (BD), jumping (JM), rotating (RT), sitting down (SD), standing up (SU), and walking (WL). Each activity is performed 200 times, resulting in 9,600 CSI time series each containing 256 samples.
Similar to gesture recognition, we employ WSST for transforming a time series into a spectrogram. To classify these activities, we utilize the same classifier as in RF-Net~\cite{RF-Net-SenSys20}. %This classifier employs metric-based meta-learning to achieve accurate activity recognition independent of changes in the environment and subjects.

\vspace{.5ex}
All evaluations focus on demonstrating \name's capability of multi-person sensing with commodity Wi-Fi devices; they, by no means, aim to show competitive performance against existing single-person monitoring systems. 
Instead, our objective is to validate \name's physical separability and quantify its benefits over non-near-field 
sensing.
The comparisons between them are done by contrasting the sensing accuracy results of the former for an arbitrary subject against those of the latter.
Our experiments have strictly followed the IRB of our institute.

%% file: 5_evaluations.tex
In this section, we begin with two micro-benchmark studies, verifying the effectiveness of SRA and further testing the differences in sensing via BFI vs CSI.
This is then followed by the three case studies specified in Section~\ref{ssec:setup}.

\vspace{-1em}
\subsection{Micro-benchmark Studies}

\subsubsection{Effectiveness of Sparse Recovery}
\begin{figure}[t] 
    \vspace{-1ex}
    \setlength\abovecaptionskip{0pt}
     \setlength\subfigcapskip{-2pt}
    \centering
	\subfigure[MSE Loss.]{
        \centering
		\label{fig: sparse recover mse loss}
		\includegraphics[width=0.42\linewidth]{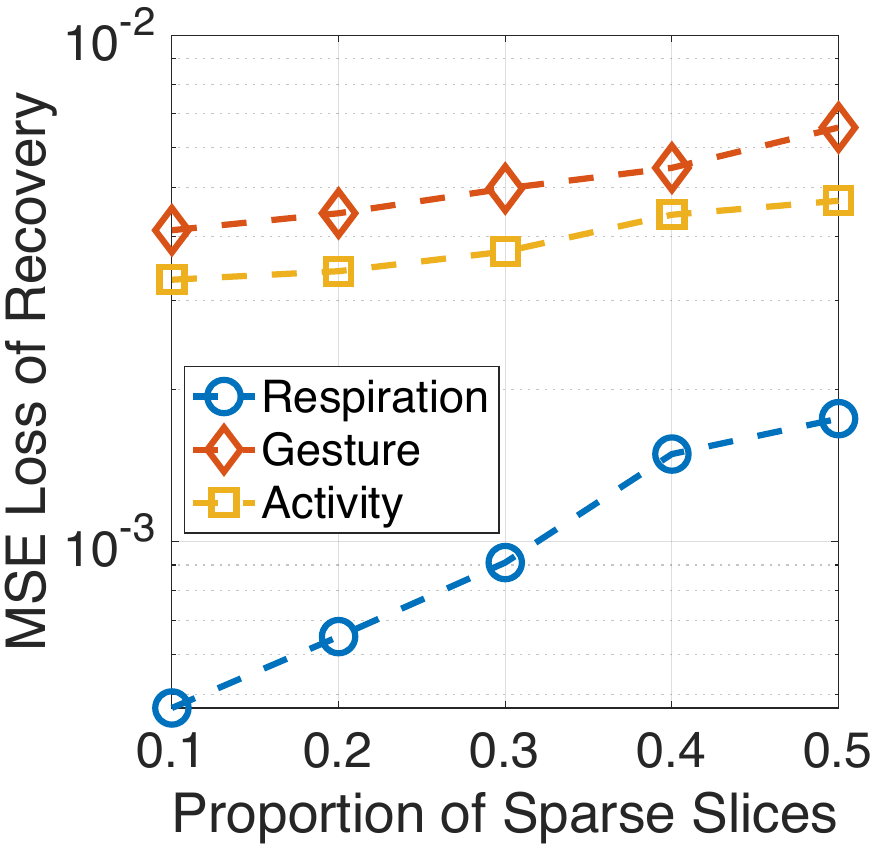}
		\vspace{-2ex}
	}
	\hfill
	\subfigure[\rev{Illustration of sparse recovery}.]{
        \centering
		\label{fig: sparse recover illustration}
		\includegraphics[width=0.52\linewidth]{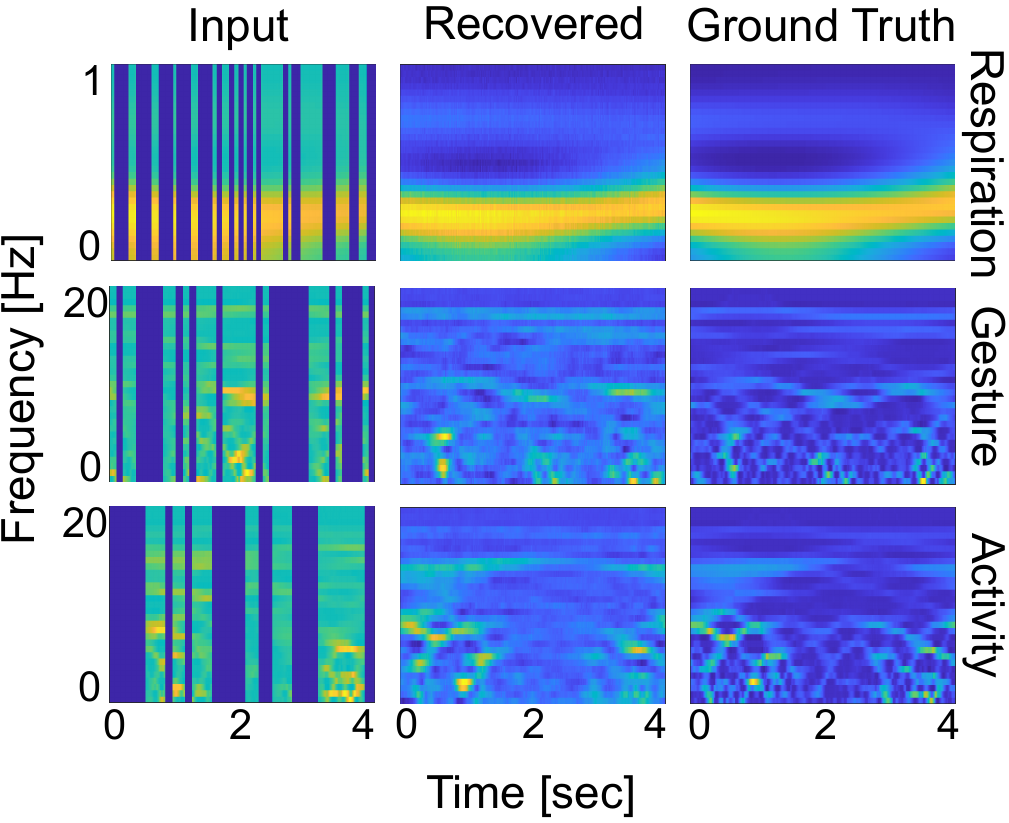}
		\vspace{-2ex}
	}
    \caption{Performance of sparse recovery.}
    \label{fig: sparse recovery}
    \vspace{-1em}
\end{figure}

To demonstrate the effectiveness of SRA, we collect CSI time series for one subject performing respiration, gesture, and activity. 
We use the MSE loss between the recovered and ground truth spectrograms as in Eqn.~\eqref{opt: training} to evaluate the performance of SRA. Figure~\ref{fig: sparse recover mse loss} displays how the MSE losses of the recovery for all three categories vary with the amount of missing slices, clearly showing the MSE losses for respiration, gesture, and activity as below $2\times10^{-3}$, $5\times10^{-3}$, and $7\times10^{-3}$, respectively. Given the normalized spectrogram data, these resulting MSE values are sufficiently low to indicate successful recovery, thus validating the effectiveness of SRA.
We also provide, in Figure~\ref{fig: sparse recover illustration}, examples of recovered spectrograms. 
It is evident that SRA successfully recovers a significant portion of the input spectrogram, albeit miss a few minor details.

Upon further inspection, it is noticeable that gestures induce the largest MSE loss in recovery. This is likely because the hands of the subject, when compared with the subject's body, are closer to the UE, making the sensing results more sensitive to hand movements.
The higher sensitivity to gestures introduces more complicated time-frequency patterns to the input spectrogram, naturally lowering the accuracy of the sparse recovery. On the contrary, respiration induces the least MSE loss because of its relatively stable and periodic style, which results in more regular patterns in the spectrogram and hence facilitates sparse recovery.

\subsubsection{Comparison between CSI and BFI}
Since UL-CSI and DL-CSI are symmetric, we refrain from comparing them but rather combine their outcomes and analysis in the following. 
To further analyze the brief observations made in Section~\ref{ssec:bfi_analysis}, we perform sensing on a subject carrying out %various movements, including respiration, gestures (raising hand), and activities (standing up and sitting down) multiple times. 
Since the time-domain results are consistent with those shown in Figure~\ref{fig: bfi csi compare time}, we do not show such results again for brevity.

\begin{figure}[b] 
    \vspace{-1.5ex}
    \setlength\subfigcapskip{-2pt}
    \setlength\abovecaptionskip{3pt}
    \centering
    \hspace{-0.36em}
	\subfigure[Respiration.]{
        %\centering
		\label{fig: bfi csi compare box re}
		\includegraphics[height=6.1em]{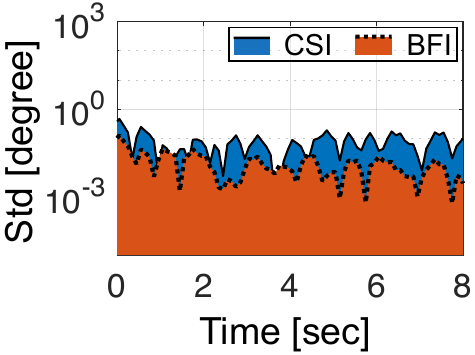}
		\vspace{-2ex}
	}
	%\hfill
        \hspace{-0.7em}
	\subfigure[Gesture.]{
        %\centering
		\label{fig: bfi csi compare box ge}
		\includegraphics[height=6.1em]{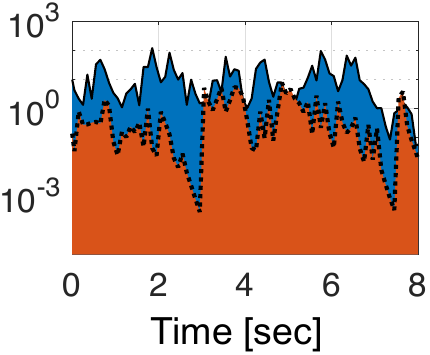}
		\vspace{-2ex}
	}
        %\hfill
        \hspace{-0.7em}
 	\subfigure[Activity.]{
        %\centering
		\label{fig: bfi csi compare box ac}
		\includegraphics[height=6.1em]{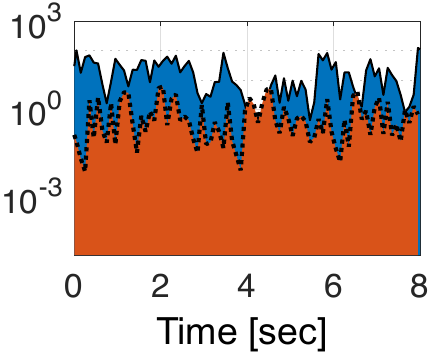}
		\vspace{-2ex}
	}
    \\[-.2em] \vspace{-2.ex}
	\subfigure[Respiration.]{
        \centering
		\label{fig: bfi csi compare time re}
		\includegraphics[height=6.3em]{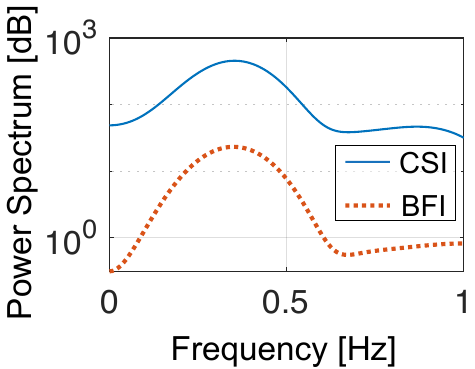}
		\vspace{-2ex}
	}
	\hspace{-0.7em}
	\subfigure[Gesture.]{
        \centering
		\label{fig: bfi csi compare time ge}
		\includegraphics[height=6.3em]{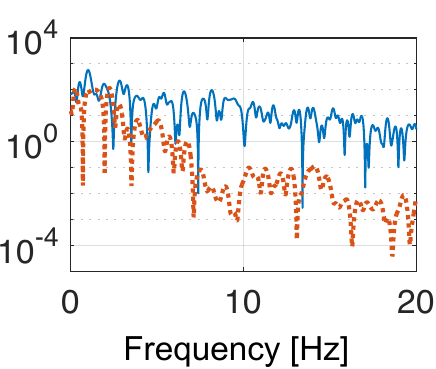}
		\vspace{-2ex}
	}
        % \hfill
        \hspace{-0.7em}
 	\subfigure[Activity.]{
        \centering
		\label{fig: bfi csi compare time ac}
		\includegraphics[height=6.3em]{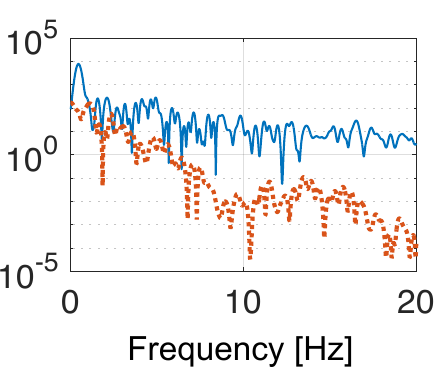}
		\vspace{-2ex}
	}
    \caption{Comparing CSI and BFI in terms of standard deviations (a)-(c) and power spectrum (d)-(f).}
    \label{fig: bfi csi compare std}
    \vspace{-1ex}
\end{figure}
\setcounter{figure}{13}
\begin{figure*}[b]
    \vspace{-1em}
    \setlength\abovecaptionskip{0pt}
     \setlength\subfigcapskip{-2pt}
    % \vspace{-1ex}
	   % \captionsetup[subfigure]{justification=centering}
		\centering
    	\subfigure[Subject A.]{
            \centering
    		\label{fig: temp-spec a}
    		\includegraphics[height=6.2em]{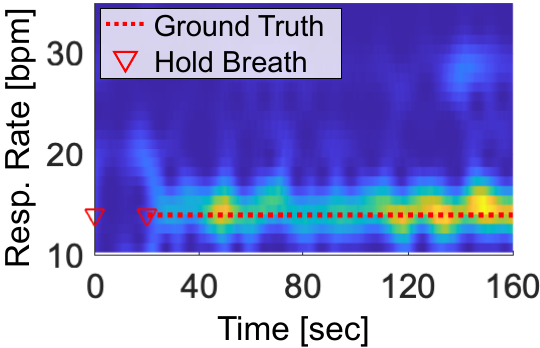}
    		\vspace{-4ex}
    	}
    	\subfigure[Subject B.]{
            \centering
    		\label{fig: temp-spec b}
    		\includegraphics[height=6.2em]{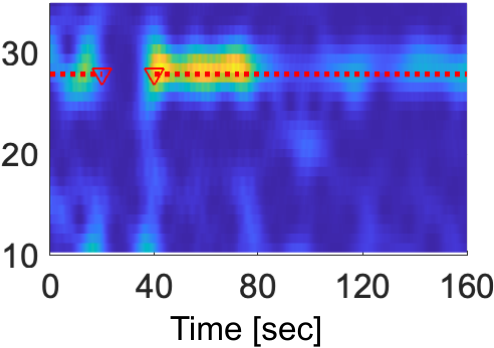}
    		\vspace{-4ex}
    	}
    	\subfigure[Subject C.]{
            \centering
    		\label{fig: temp-spec c}
    		\includegraphics[height=6.2em]{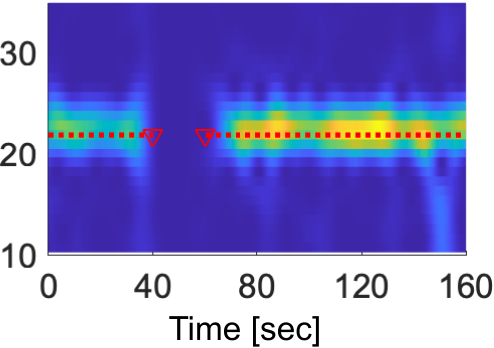}
    		\vspace{-4ex}
    	}
    	\subfigure[Subject D.]{
            \centering
    		\label{fig: temp-spec d}
    		\includegraphics[height=6.2em]{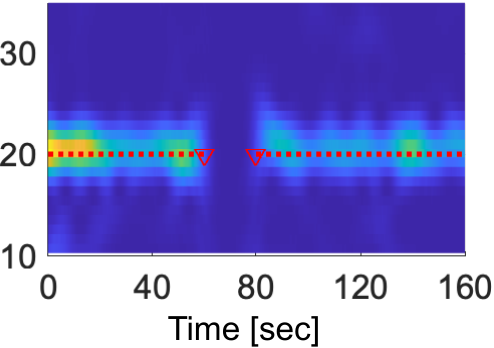}
    		\vspace{-4ex}
    	}
    	\subfigure[Subject E.]{
            \centering
    		\label{fig: temp-spec e}
    		\includegraphics[height=6.2em]{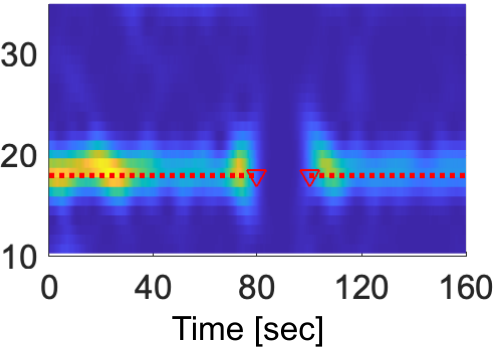}
    		\vspace{-4ex}
    	}
    	\\[-.5em] \vspace{-1.6ex}
    	\subfigure[Subject F.]{
            \centering
    		\label{fig: temp-spec f}
    		\includegraphics[height=6.2em]{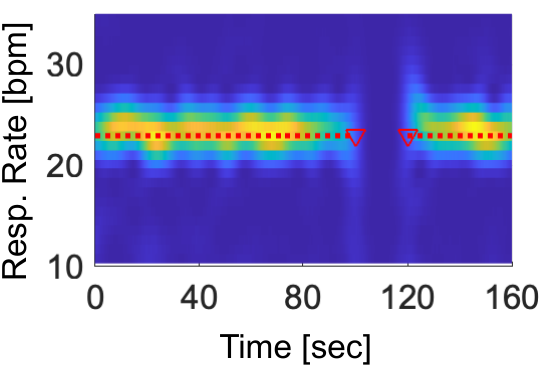}
    		\vspace{-4ex}
    	}
    	\subfigure[Subject G.]{
            \centering
    		\label{fig: temp-spec g}
    		\includegraphics[height=6.2em]{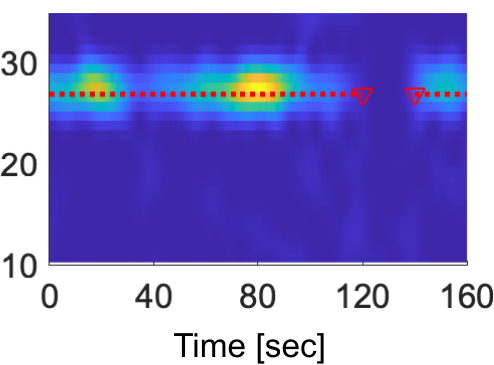}
    		\vspace{-4ex}
    	}
    	\subfigure[Subject H.]{
            \centering
    		\label{fig: temp-spec h}
    		\includegraphics[height=6.2em]{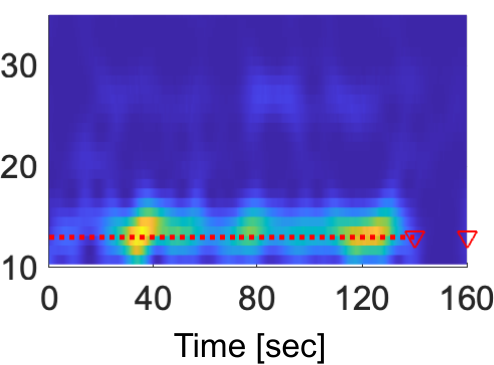}
    		\vspace{-4ex}
    	}
    	\subfigure[Baseline.]{
            \centering
    		\label{fig: temp-spec baseline}
    		\includegraphics[height=6.2em]{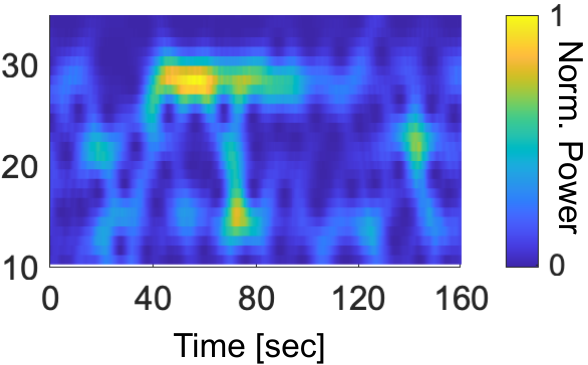}
    		\vspace{-4ex}
    	}
    	\subfigure[Entropy Measure.]{
            \centering
    		\label{fig: temp-spec entropy}
    		\includegraphics[height=6.2em]{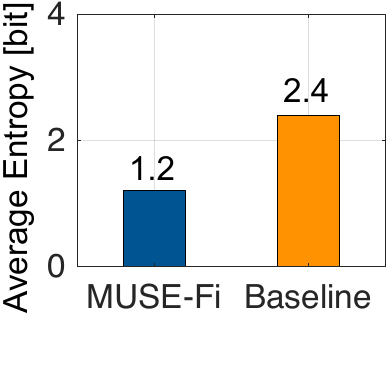}
    		\vspace{-4ex}
    	}
		\caption{Comparison and analysis on \name and the baseline, in terms of the respiration spectrograms.}
		\label{fig: temp-spec compare}
\end{figure*}
We further investigate the fluctuations of the BFI and CSI signals by calculating the standard deviations of the detrended signals over periods of 0.1 seconds, as shown in Figures~\ref{fig: bfi csi compare box re} to \ref{fig: bfi csi compare box ac}.
The figure reveals that BFI signals are more stable than CSI signals, but this stability comes at the cost of reduced sensitivity.
We also examine the power spectrum of the CSI and BFI in Figures~\ref{fig: bfi csi compare time re} to \ref{fig: bfi csi compare time ac}. 
One may readily observe that while CSI preserves the respiration signal and presents a smooth spectrum, it is too sensitive to rapid and large-scale motions and results in excessive power in high frequencies for gesture and activity. 
In comparison, BFI effectively suppresses high-frequency components, while its response to low-frequency subtle movements (e.g., respiration) is less pronounced.

To explain these phenomena, it is worth noting that the phase of CSI is directly related to relative displacement, making it sensitive to small-scale movements (e.g., respiration). However, large-scale movements cause abrupt phase changes that cannot be captured by insufficient sampling, resulting in irregularities in the CSI signal. As explained in Section~\ref{ssec:bfi_analysis}, the BFI-based sensing strategy only captures the relative directional changes from the subject to the AP: if one deems the conversion from CSI to BFI as ``low-pass'' filtering, it would be natural to expect \rev{fewer} variations but also lowered strength in the resulting signals. This property is particularly beneficial for a future study on subjects carrying smartphones on \rev{their} bodies for continuous vital signs monitoring~\cite{MoVi-Fi-MobiCom21,zheng2021more}, as BFI sensing may filter out body movement interference.

\vspace{-.8em}
\subsection{Case-I: Respiration Monitoring} \label{ssec:resp}
We conduct experiments to monitor multi-person respiration using the setup described in Section~\ref{ssec:setup}. After obtaining sparse recovery results,
we use a 3-layer CNN to extract the respiration rate and measure the respiration rate error as $|R_{\text{E}} -R_{\text{A}}|$, where $R_{\text{E}}$ is the estimated respiration rate and $R_{\text{A}}$ is the actual respiration rate. 
Figure~\ref{fig: resp waveform} showcases the respiration waveforms obtained by \name and the baseline method. One may clearly observe that \name recovers the respiration waveforms effectively, whereas the baseline method only captures a noisy signal mixture contributed by multiple subjects. We further assess the accuracy of respiration rate estimation of both \name and the baseline in Figure~\ref{fig: resp error}; the results reveal that \name achieves accurate respiration monitoring with both median and mean respiration rate errors less than 1~\!bpm. In contrast, the baseline exhibits a median and mean respiration rate error of 7 and 8~\!bpm, respectively, making it almost useless in multi-person scenarios.
\setcounter{figure}{12}
\begin{figure}[t] 
    \vspace{-1ex}
    \setlength\abovecaptionskip{0pt}
    \centering
	\subfigure[Respiratory waveform.]{
        \centering
		\label{fig: resp waveform}
		\includegraphics[height=6.8em]{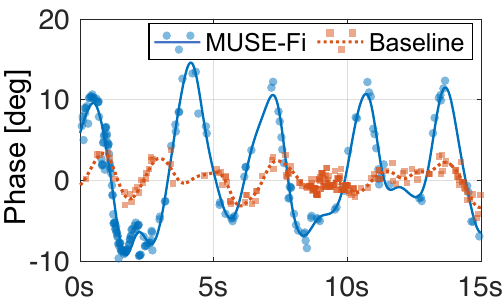}
		\vspace{-4.5ex}
	}
	\hspace{-0.5em}
	\subfigure[Sensing error.]{
        \centering
		\label{fig: resp error}
		\includegraphics[height=6.8em]{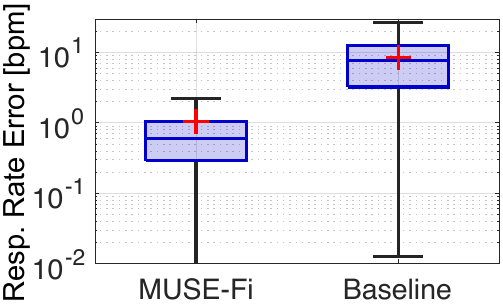}
		\vspace{-4.5ex}
	}
    \caption{Comparison between \name and the baseline in terms of respiration sensing.}
    \label{fig: respiration monitoring}
    \vspace{-2em}
\end{figure}

\setcounter{figure}{14}
\begin{figure*}[t]
    %\vspace{-2ex}
    \setlength\abovecaptionskip{3pt}
     \setlength\subfigcapskip{-2pt}
    \centering
    \subfigure[Summary.]{
        \centering
        \label{fig:ge_ac_summary}
        \includegraphics[height=8.6em]{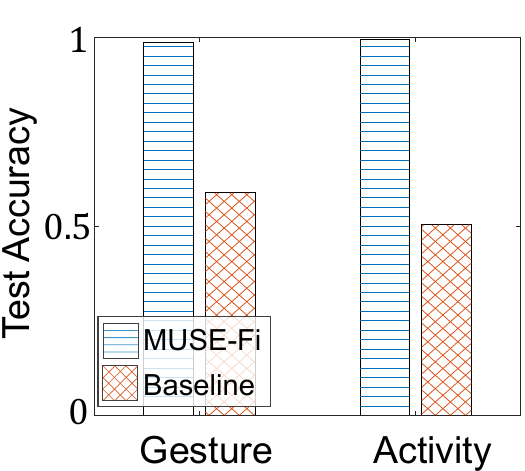}
        \vspace{-2ex}
    }
    \hspace{-.6em}
    \subfigure[Gesture (\name ).]{
        \centering
        \label{fig:ge_muse}
        \includegraphics[height=8.3em]{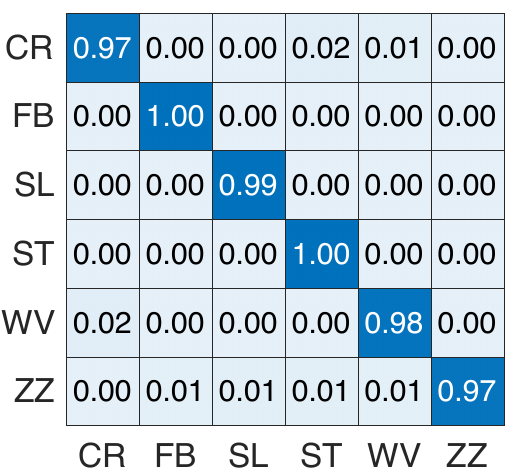}
        \vspace{-2ex}
    }
        \hspace{-.6em}
    \subfigure[Gesture (baseline).]{
        \centering
        \label{fig:ge_baseline}
        \includegraphics[height=8.3em]{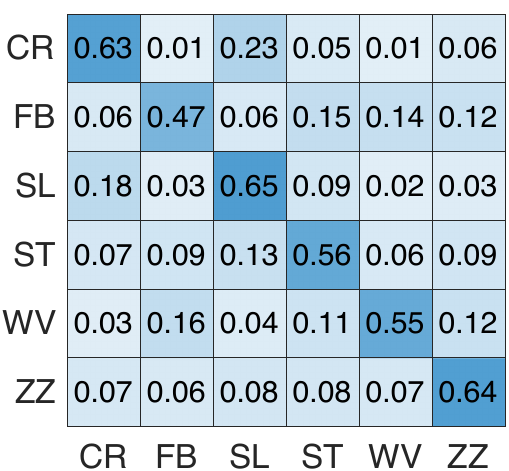}
        \vspace{-2ex}
    }
        \hspace{-.6em}
    \subfigure[Activity (\name ).]{
        \centering
        \label{fig:ac_muse}
        \includegraphics[height=8.3em]{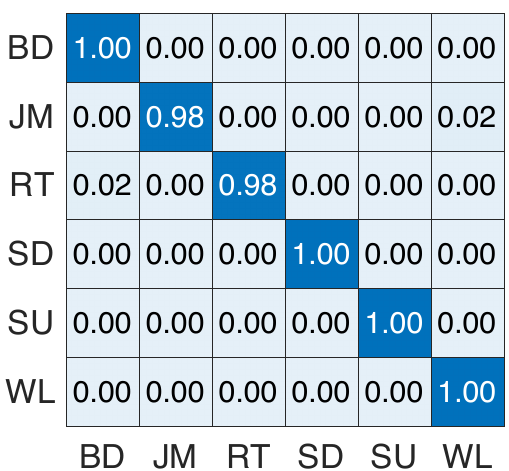}
        \vspace{-2ex}
    }
        \hspace{-.6em}
    \subfigure[Activity (baseline).]{
        \centering
        \label{fig:ac_baseline}
        \includegraphics[height=8.3em]{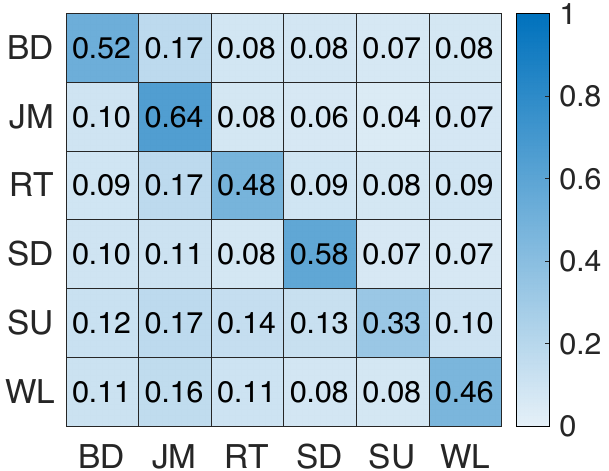}
        \vspace{-2ex}
    }
    \caption{Comparison between \name and baselines for gesture and activity recognition.}
    \label{fig:ge_ac_comp}
    \vspace{-2ex}
\end{figure*}

To further understand the performance difference, we present the spectrograms of \name and the baseline method in Figures~\ref{fig: temp-spec a}-\ref{fig: temp-spec h} and~\ref{fig: temp-spec baseline}, respectively. 
We can observe that \name recovers a clear signal around the ground truth frequency thanks to the near-field domination. 
Moreover, we let the subjects sequentially hold their breath for 20~\!s, and the correspondence between the breath-holding periods and signal interruption (whose boundary is denoted by two triangular markers) on the spectrograms firmly proves that the respiration signals from different subjects are well separated. 
On the contrary, the baseline method fails to distinguish respiration from multiple subjects, resulting in a noisy spectrogram where no accurate respiration rate can be obtained. 
We further employ the average spectral entropy~\cite{shannon2001mathematical} of the normalized spectrogram to measure the residual uncertainty in determining respiration rate. 
Specifically, Figure~\ref{fig: temp-spec entropy} indicates \rev{a} $2.4$~\!bit entropy for the baseline, much higher than the $1.2$~\!bit entropy of \name. 
Intuitively, the variety of potential respiration rates represented by the spectrogram increases exponentially with its spectral entropy.
Therefore, this halved entropy value implies that the respiration rate decision of \name can be much more precise than that of the baseline, thus explaining our result in Figure~\ref{fig: resp error}.
All these results provide conclusive evidence of the efficacy of \name in respiration monitoring.

\vspace{-.5ex}
\subsection{Case-II: Gesture Detection} \label{ssec:gesture}
We also conduct experiments on gesture detection, and summarize the statistics in Figure~\ref{fig:ge_ac_summary}. 
One may readily observe that \name achieves a mean test accuracy of more than $98$\%, while that of the baseline is only $57$\%. 
We further inspect the confusion matrices of \name and the baseline in Figure\rev{s}~\ref{fig:ge_muse} and~\ref{fig:ge_baseline}, respectively. 
The confusion matrices reveal that \name can correctly classify most gestures, while the baseline often confuses one gesture with others. 
Specifically, the circle~(CR) and slide~(SL) gestures are the most \rev{confusing} pair \rev{for} the baseline, as they both involve moving one's hands smoothly over the phone, which may appear similar to the baseline in the far field, but are readily differentiable by the near-field \name. 
The baseline’s inferior performance can also be attributed to its inability to disentangle signals from interference caused by the moving hands of multiple people, eventually causing its unacceptable detection behavior (i.e., $39$\% lower than \name). 
These results evidently confirm \name's effectiveness in resolving multi-person gesture detection for real-world applications.

\vspace{-.5ex}
\subsection{Case-III: Activity Recognition} \label{ssec:har}
We further conduct experiments on activity recognition, with statistics summarized in Figure~\ref{fig:ge_ac_summary}; \name's mean accuracy of more than $98$\% is doubled of the baseline's that drops by 8\% compared with the gesture detection task. 
This performance degradation is attributed to the greater interference induced by large-scale and rapid human activities. 
We further inspect the confusion matrices of \name and the baseline in Figure\rev{s}~\ref{fig:ac_muse} and~\ref{fig:ac_baseline}, respectively. 
One may readily observe that the accuracy for all $6$ activities is above $0.98$ for \name, while the baseline's accuracy is all less than $0.52$ (with SU bearing the worst accuracy of $0.33$). 
The results from all three case studies have successfully demonstrated a great potential to realize a long-standing vision
for Wi-Fi human sensing: multiple people sitting around a table (e.g., holding a meeting), while leveraging contactless sensing to accomplish diversified tasks with the support of their respective smartphones and only one Wi-Fi AP.

\vspace{-.5em}
\subsection{Extended Experiments and Discussions} \label{ssec: extend experi}
\input{7_discussion.tex}

%% file: 7_discussion.tex
\begin{figure}[b]
    \vspace{-1em}
    \setlength\abovecaptionskip{0pt}
     \setlength\subfigcapskip{-2pt}
    \centering
	\subfigure[Gesture]{
		\label{fig: nlos gesture a}
		\includegraphics[width=0.465\linewidth]{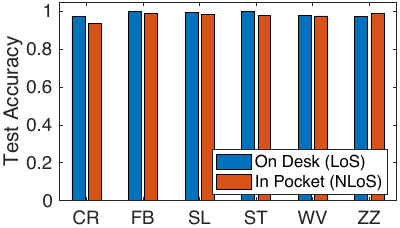}
		\vspace{-2ex}
	}
	\hfill
	\subfigure[Activity.]{
		\label{fig: nlos activity b}
		\includegraphics[width=0.47\linewidth]{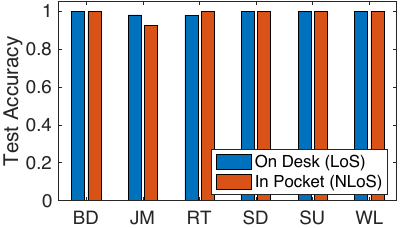}
		\vspace{-2ex}
	}
    \caption{\frev{Comparison between the sensing accuracy of \name\ under LoS and NLoS conditions.}}
    \label{fig: nlos gesture and activity}
    \vspace{-1ex}
\end{figure}

To prove the generalizability of \name, we evaluate it in another practical scenario, where the subjects carry their smartphones inside their pockets, hence the LoS paths between the AP and UEs are blocked. 
Here we focus on the gesture detection and activity recognition tasks, given their relevance to enabling the computer-human interfacing for XR applications.
In Figure~\ref{fig: nlos gesture and activity}, we compare the sensing accuracy of \name\ in two scenarios: 1) the on-desk scenario with LoS condition as in Figure~\ref{fig: respiration scene layout}, and 2) the in-pocket scenario with non-LoS~(NLoS) condition, where the sensing accuracy of \name\ are shown to be similar and higher than $92\%$ for the both scenarios.
This is because the Wi-Fi signals can diffract and bypass the boundary of body, while the condition for near-field domination effect still holds.

\frev{Based on the above case studies and extended experiments, we make the following discussions on \name's generalizability, potential applications, and key factors.}

\vspace{-.5em}
\frev{\paragraph{Generalizability} 
\name\ is capable of generalizing beyond current experimental setup because, firstly, environment dynamics have a small impact on \name\ due to the near-field domination effect; and secondly, environment layout changes typically manifests as additional biases to the CSI and can be removed during the normalization of SRA.}

\vspace{-.5em}
\frev{\paragraph{Potential Applications}
With the physical separability guaranteed by the near-field domination effect, \name\ is scalable to ubiquitous Wi-Fi networks in daily life and can provide solutions for XR.
Based on respiration monitoring results in Section~\ref{ssec:resp} \name\ can track subtle motions of subjects near distributed UEs, which can extend to AR/MR solutions for visualizing intrusion, vital signs of human, and operation status of machines.}
\frev{Besides, the results in Sections~\ref{ssec:gesture},~\ref{ssec:har}, and~\ref{ssec: extend experi} indicate its capability to accurately recognize individuals’ gestures and activities.
This means \name\ enables the sensing functionality of gesture detection and activity recognition to be integrated into Wi-Fi modules naturally carried by each individual, potentially reducing the cost, weight, and power consumption of VR/MR headsets for computer-human interaction.}

\vspace{-.5em}
\frev{\paragraph{Key Factor Analysis}
The near-field domination effect assumes the existence of LoS UE-AP paths, yet \name\ is also robust to the LoS blockage by the subject's body as shown in Figure~\ref{fig: nlos gesture and activity}.
Therefore, \name\ is effective in common practice where APs are located above the subjects.
If LoS paths are blocked, and the signals travel via NLoS paths involving reflection in the environment, then the near-field domination condition can be naturally extended to the shortest NLoS paths, provided that these paths are clear of environment dynamics (e.g., other irrelevant motions) to avoid injecting interference directly into the received signals.}
\frev{In addition, \name\ efficiently handles the sparsity of realistic data traffic for online videos and meetings etc, while the highly sparse data traffic for idle UEs may be beyond recovery and lead to invalid sensing results.}
Finally, sensing security~\cite{WiKI-Eve-CCS23, mimoCrypt-SP24} and co-existing with other co-channel communication systems~\cite{muzi, favor, art} should be handled in the future.

%% file: 6_related_work.tex
Our work is closely related to contactless human sensing~\needrev{\cite{LiFS-MobiCom16,WiAG-MobiSys17,Widar2-MobiSys18,Widar3-MobiSys19,WiPose-MobiCom20,Wang2022Placement,VitalSign-MobiHoc15,MultiSense-UbiComp20,adib2015smart,MoVi-Fi-MobiCom21,RF-Net-SenSys20,song2020spirosonic, xu2019breathlisener, yun2017strata, pegoraro2022sparcs}} that has experienced significant advancements in the past decade. Of all these solutions, Wi-Fi human sensing~\cite{LiFS-MobiCom16,WiAG-MobiSys17,Widar2-MobiSys18,Widar3-MobiSys19,WiPose-MobiCom20,Wang2022Placement,VitalSign-MobiHoc15,MultiSense-UbiComp20,azizyan2009surroundsense} gains particular popularity because it is compatible with existing communication hardware, thus not incurring additional deployment cost. Specifically, Wi-Fi human sensing exploits CSI~\cite{CSI-CCR11} retrieved from the received signals of Wi-Fi communications. The ubiquity of Wi-Fi infrastructure (e.g., Wi-Fi APs, laptops, and smartphones) has led to a multitude of research studies for various applications, including vital sign monitoring~\cite{VitalSign-MobiHoc15,Wang2022Placement}, gesture detection~\cite{Widar3-MobiSys19, WiAG-MobiSys17}, activity recognition~\cite{WiPose-MobiCom20, RF-Net-SenSys20}, localization~\cite{SpotFi-SIGCOMM15,soltanaghaei2018multipath}, and motion tracking~\cite{LiFS-MobiCom16,sun2015widraw}, which, at a higher level, are driving the vision of smart home~\cite{adib2015smart} and digital healthcare~\cite{ge2022contactless, tan2018exploiting}.

While the above proposals mainly focus on single-person scenarios due to the limited range resolution of existing Wi-Fi technology, recent research has explored the use of next-generation Wi-Fi technologies to overcome this challenge. ViMo~\cite{Wang2021ViMo} uses $60$~\!GHz 802.11ad devices~\cite{qualcomm} with high bandwidth and a 32-element phased array to emulate a radar. Similarly, mmTrack~\cite{wu2020mmtrack} employs the same 802.11ad device for multi-person localization. However, the limited adoption and high cost of 802.11ad devices may hinder the goal of multi-person monitoring. It should be noted that \name shares part of its name with MUSE~\cite{MUSE-MobiCom16}, but these two systems bear distinct objectives: whereas MUSE focuses on communication scheduling under MU-MIMO, \name targets multi-person sensing.

Other techniques than future Wi-Fi hardware may also help enhance human sensing. 
Widar2.0~\cite{Widar2-MobiSys18} provides partial support for multi-person sensing by using multiple antennas to improve spatial resolution. 
Karanam \textit{et al}.~\cite{karanam2019tracking} use the magnitude measurements from an array of receivers to perform multi-person tracking. 
Lan \textit{et al}.~\cite{lan2020wireless} \rev{employ} metasurface antennas with varying beam patterns to perform multi-person activity recognition.  %Algorithmic approaches also constitute a significant portion of these techniques. 
Liu et al.~\cite{VitalSign-MobiHoc15} estimate multi-person respiration rates by analyzing CSI's power spectral density. PhaseBeat~\cite{Wang2017PhaseBeat} and TR-BREATH~\cite{Chen2018TRBREATH} leverage root-MUSIC~\cite{Rao1989music} to separate multi-person sensing signals, while Yang et al.~\cite{Yang2018Multi} optimize transceiver deployment using the Fresnel zone model to reduce interference, but with the requirement of accurate subject location and fixed transceiver placement.
MultiSense~\cite{MultiSense-UbiComp20} treats multi-person sensing as a blind source separation problem and uses ICA~\cite{hyvarinen2000ica} to extract waveforms. 
Last but not least, SPARCS~\cite{pegoraro2022sparcs} \rev{recovers} the micro-doppler spectrum by using intrinsic sparsity of \rev{wideband} mmWave channels.
However, it does not fit for \rev{narrowband Wi-Fi systems operating at microwave band.}

%% file: 8_conclusion.tex
Taking an important step towards ubiquitous human sensing, \name has innovated in Wi-Fi multi-person sensing by addressing the major challenge of physically separating multiple subjects. Leveraging the near-field channel variation caused by a subject in close proximity to a Wi-Fi device, \name\ \rev{has} demonstrated successful handling of multi-person sensing for respiration monitoring, gesture detection, and activity recognition. 
This success also stems from our two technical developments: i) an SRA to cope with realistic (intermittent) Wi-Fi traffic under multi-user scenarios, and ii) a study on the difference between CSI and BFI sensing.
Our extensive evaluations have evidently confirmed that \name is a cost-effective alternative to radar-based systems that often require extra deployments.
Moving forward, we believe that \name\ has significant potential to be extended into various applications, including healthcare, smart homes, and even security; we are also planning to deploy \name on larger scales so as to evaluate its performance in more diversified scenarios. 

\vspace{-.33em}
\section*{Acknowledgement}
This research is support by National Research Foundation (NRF) Future Communications Research \& Development Programme (FCP) grant FCP-NTU-RG-2022-015. % and MOE Tier 1 grant RG16/22.